\documentclass[a4paper,11pt]{article}

\usepackage{amsmath}
\usepackage{amssymb}
\usepackage{amsfonts}
\usepackage{graphicx}
\usepackage{color}

\newcommand{\lsim}{\mathrel{\lower0.55ex\hbox{$\mathchar"3218$}\mkern-14mu\raise0.55ex\hbox{$<$}}}
\newcommand{\gsim}{\mathrel{\lower0.55ex\hbox{$\mathchar"3218$}\mkern-14mu\raise0.55ex\hbox{$>$}}}

\begin{document}

\title{A moving boundary problem motivated by electric breakdown:
I. Spectrum of linear perturbations}

\author{S. Tanveer$^1$, L. Sch\"afer$^2$, F. Brau$^3$, U. Ebert$^{3}$\footnote{corresponding author} \\~\\
$^1$ Department of Mathematics, Ohio State University, 
USA,\\
$^2$ Fachbereich Physik, Universit\"at Duisburg-Essen, 
Germany,\\
$^3$ Centrum Wiskunde \& Informatica (CWI), \\P.O.Box 94079, 1090GB
Amsterdam, The Netherlands.\\
}

\date{\today}

\maketitle

\begin{abstract}
An interfacial approximation of the streamer stage in the evolution of sparks and lightning can be written
as a Laplacian growth model regularized by a `kinetic undercooling' boundary condition. We study the linear
stability of uniformly translating circles that solve the problem in two dimensions. In a space of smooth
perturbations of the circular shape, the stability operator is found to have a pure point spectrum. Except
for the eigenvalue $\lambda_0=0$ for infinitesimal translations, all eigenvalues are shown to have negative
real part. Therefore perturbations decay exponentially in time. We calculate the spectrum through a
combination of asymptotic and series evaluation. In the limit of vanishing regularization parameter, all
eigenvalues are found to approach zero in a singular fashion, and this asymptotic behavior is worked out in
detail. A consideration of the eigenfunctions indicates that a strong intermediate growth may occur for
generic initial perturbations. Both the linear and the nonlinear initial value problem are considered in a
second paper.
\end{abstract}

\newpage

\section{Introduction} \label{introduction}

The motion of interfaces in a Laplacian field is of general interest and has been a subject of intense study
over many years (see for instance the reviews \cite{Kessleretal,Pelce}). Such problems arise in many
physical contexts, such as viscous fingering in multi-phase fluid flow \cite{Tanveer86, Bensimon1986,
bens86, DeGregoriaSchwartz86, KesslerLevine86, Combescotetal86, Howison86, TanveerSaff87, Tanveer87b,
Tanveer87c, DorseyMartin, TanveerSaff88, Tanveer00, how-92}, dendritic crystal growth in the quasi-steady
small Peclet number limit \cite{DombreHakim87, EBrener1991, Xu1991, Kunkaetal}, void electromigration
\cite{ ho-70, mah-96, amar-99, Cummingsetal, kuhn-05} and a host of other phenomena such as the growth of
biological systems like bacterial colonies or corals \cite{Judith-02}.

More recently, a similar mathematical problem has been discovered in the study of `streamers'
\cite{Lozansky, DW87, Vit94, Raizer, BazRai, eber97, eber06, brau08} which occur during the initial stage of
electric breakdown and play an important role both in the natural phenomena of sparks and lightning as well
as in numerous technical applications~\cite{eber06}. Streamers are weakly ionized bodies growing into some
nonionized medium due to an externally applied electric field. This field is so strong that the drifting
electrons very efficiently create additional electron ion pairs by impact ionization, and the nonlinear
coupling between ionized body and field further increases this effect.

Models for negative streamers in simple gases like nitrogen or argon are based on a set of partial
differential equations for the densities of electrons and of positive ions coupled to the electric
field~\cite{DW87,Vit94,Raizer,BazRai,eber97,eber06}. Analysis and numerical solutions of these equations
reveal that in the front part of the streamer, a thin surface charge layer develops where the electron
density strongly exceeds the ion density. Therefore the electric field ${\bf E}$ varies strongly when
crossing this layer. Right before the layer, it is enhanced, but in the interior of the streamer, it is
screened to such a low level that impact ionization is suppressed and the electron current transporting
charge from the interior to the surface charge layer is small. Consequently we may take the interior as
being essentially passive, and the growth of the streamer is governed by the surface charge layer which is
driven by the strong local field.

If the external field is very strong, the thickness $\ell$ of the surface charge layer can become small
compared to the typical diameter $2 R$ of the streamer \cite{brau08}. This suggests modeling this layer as
an interface separating the ionized interior from the nonionized exterior region. In this model the
variation of the potential $\varphi$ across the surface charge layer is replaced by a discontinuity on the
interface. Since the interior is considered passive, only the limiting value $\varphi^+$ reached by
approaching the interface from the outside is relevant for the dynamical evolution, and analysis of results
of the PDE-model suggests \cite{brau08,meul05,eber07} that with an appropriate gauge, $\varphi^+$ is coupled
to the limiting value ${\bf E}^+$ of the electric field by the boundary condition
\begin{equation} \label{Achat}
\varphi^+ = - \ell {\bf n} \cdot {\bf E}^+ \,.
\end{equation}
Here ${\bf n}$ is the outward normal on the interface, $\ell$ is the regularization length corresponding to
the interface thickness, and ${\bf E}^{+}=-\nabla \varphi^+$, where the $^{+}$ again indicates the limit of
approaching the interface from the outside. In the context of dendritic crystal growth, this boundary
condition is equivalent to including the kinetic undercooling effect\footnote{Under most natural conditions
of crystal growth, kinetic undercooling is important in a limit when the Peclet number is not small enough
to justify a Laplacian field approximation; nonetheless, there have been some studies of steady Laplacian
crystal growth with kinetic undercooling effects only \cite{Chapman-King}.} while excluding the usual
Gibbs-Thompson surface energy correction to the melting temperature.

As the motion of the interface is caused by the drift of the electrons in the local electric field ${\bf
v}=-{\bf E}$, the interface moves with normal velocity
\begin{equation} \label{Amazonit}
v_n = -{\bf n}\cdot{\bf E}^+,
\end{equation}
and outside the streamer the potential obeys the Laplace equation
\begin{equation} \label{Amethyst}
\Delta \varphi = 0 \, .
\end{equation}

We discuss the problem defined by Eqs.~(\ref{Achat})-(\ref{Amethyst}) in infinite two-dimensional space,
with the electric field becoming constant, ${\bf E} =-\nabla \phi \to {\bf E}_\infty$, far from the
streamer; such a condition is realized frequently in atmospheric discharges, e.g., inside thunderclouds.
This far-field condition on $\nabla \phi$ is different from the usual source/sink condition in viscous
fingering in the absence of side-walls, or undercooling specification in the quasi-steady low Peclet number
crystal growth problem which have been extensively studied. However, moving bubbles and fingers in a long
Hele-Shaw channel are indeed subjected to this type of far-field condition. (For the discussion of the
streamer problem equivalent to the Saffman-Taylor finger, we refer to~\cite{luqu08}.) Further, in the
crystal growth or directional solidification problem, a systematic inner-outer analysis for small Peclet
number \cite{DombreHakim87,Kunkaetal} shows that this is an appropriate condition at $\infty$ for the
"inner" problem.

A simple steady solution to the streamer equations given above is a circle translating with constant
velocity determined by ${\bf E}_\infty$~\cite{meul05,eber07}. Though such circles differ from proper
streamers, which are growing channels of ionized matter~\cite{eber06,luqu08,arra02,caro06}, their front half
closely resembles the head of the streamer where the growth takes place. It is therefore a question of
physical interest whether or not translating circular solutions are stable to small perturbations and this
is the subject of the present investigation.

The relevance of this analysis for more realistic streamer shapes is supported by results found in another
physical context. Steadily translating circles also arise in viscous fingering in a Hele-Shaw cell when
surface tension is included (instead of kinetic undercooling) in the limit when the bubble is small compared
to the cell-dimensions \cite{Tanveer86}. The linear stability of these bubbles, including larger
non-circular steadily translating bubbles, has been studied before both for one and two fluids
\cite{TanveerSaff87,TanveerSaff88} and the results largely mimic those obtained for a finger, though the
latter calculations are mathematically much more involved.

It has been known for a while that in the absence of any regularization, such as surface tension or kinetic
undercooling, the initial value problem in a Laplacian field is ill-posed \cite{Howison86} in any norm that
is physically relevant to describing interfacial features. This is reflected in the instability of any
steady shape, when the growth rates increase with the wave numbers of the disturbances. Ill-posedness makes
idealized model predictions sometimes physically irrelevant (see \cite{Tanveer00} for a thorough discussion)
and regularization becomes essential.

Considering a {\em planar} front, one finds that regularization does not remove the instability against
fluctuations of small wave number. However, for large wave numbers, linear stability analysis exhibits a
basic difference between surface tension and kinetic undercooling. All large wave number components of a
disturbance decay with surface tension regularization, while for kinetic undercooling the growth rate
saturates to a constant that scales as $\ell^{-1}$~\cite{how-92}; for streamers, such a saturating
dispersion relation is derived and discussed in~\cite{arr-04,derks-08}.

For {\em curved} fronts, one can pose the question: how does curvature stabilize, if at all, a disturbance
whose wavenumber is in the unstable regime for a flat interface, either with surface tension or with kinetic
undercooling regularization? With surface tension regularization, some answers are available in the existing
literature. Arguments have been presented \cite{Bensimon1986,DeGregoriaSchwartz86} that suggest that a
localized wave packet with wave numbers in the unstable regime\footnote{Localized disturbances refer to
those with wavelengths far smaller than the typical radius of curvature of the steady shape. These can be
unstable only if the regularization parameter is sufficiently small.} advects along the front as it grows;
once it reaches the side of the front where the local normal velocity is zero, the disturbance stops
growing. If the steady shape is closed, as it is for a circle, the continued advection of the wave-packet
towards the receding parts of the interface will cause the disturbance to decay eventually\footnote{Surface
tension causes localized disturbances to decay as they advect to the sides even when an interface is not
closed but becomes parallel to the direction of motion as is the case for a finger in a Hele-Shaw cell.
However, no decay is expected for kinetic regularization. This is where a closed interface is different.}.
If regularization is small, there is a large transient growth. Unless the disturbance amplitude is smaller
than a threshold that shrinks to zero with regularization, the transient exponential growth causes the
interface to enter a nonlinear regime that can destabilize the steady front, even when it is predicted to be
linearly stable. Analysis of approximate equations, supported by numerical calculations of the full
equations support the above scenario. Similar stabilization should occur for the kinetic undercooling
boundary condition as well, though we are not aware of any explicit study affirming this expectation.

Note, however, that stabilization of localized wave packets does not rule out instability to long ranged
disturbances. A formal asympotic study for small nonzero surface tension \cite{Tanveer87c, TanveerSaff88} as
well as numerical studies \cite{KesslerLevine86} reveal that surface tension stabilizes precisely one branch
of steady solutions for fingers and bubbles in a Hele-Shaw cell. Similar results follow for a needle crystal
\cite{EBrener1991} though in the latter case, convective instability of wave packets caused by significant
normal speed along the parabolic front is believed to cause dendritic structures \cite{Kessleretal}. These
conclusions have been challenged at times by alternate scenarios (see for instance \cite{Xu1991}) that are
based on formal calculations, but with different implicit assumptions. Such controversy affirms the need for
more rigorous mathematical studies of the stability problem, even if it is for relatively simple shapes such
as the circle in the present study.

For the kinetic undercooling boundary condition, relying merely on a numerical study to understand the long
time behavior is fraught with difficulties. One finds a collapsing spatial scale for large time at the rear
of the circle. Analytically, this is found for $\epsilon=\ell/R=1$ in~\cite{meul05,eber07}\footnote{We
recall that $R$ is a measure of the size of the streamer. The precise definition is given in
Eq.~(\ref{Antimonit}) below.}. As will be argued in the present and the companion paper, the occurrence of
this collapsing scale is a general feature for any $\epsilon > 0$. This means that as $t \rightarrow
\infty$, one must resolve progressively finer scales near the back of the bubble. Further, calculations for
small $\epsilon$ require resolving a large number of transiently growing modes. All this underscores the
need of some progress on the analytical side.

The present paper, which is part I of a two-paper sequence, is devoted to the spectral properties of the
linear stability operator, associated with infinitesimal perturbations of a circle. In part
II~\cite{partII}, we will consider the initial value problem, presenting analytical and numerical results on
the evolution of both infinitesimal and finite perturbations.

The present paper is organized as follows. In Section \ref{reformulation} we reformulate the problem defined
by Eqs.~(\ref{Achat})--(\ref{Amethyst}) by standard conformal mapping, and we present the PDE governing the
time evolution of infinitesimal perturbations of the circle. This material has been presented before
in~\cite{meul05,eber07}, where also the general solution of the PDE in the case $\epsilon = 1$ has been
discussed in detail. The explicit solution found for $\epsilon = 1$ shows that outside any fixed
neighborhood of the rear of the bubble, the long-term behavior of infinitesimal perturbations is described
by $ \sum_{n=0}^\infty e^{\lambda_n t} \beta_{\lambda_n} $, where $\lambda_n $ is the $n$th eigenvalue
(ordered according to absolute value) of the linear stability operator and $\beta_{\lambda_n}$ is the
corresponding eigenfunction.

We then study this eigenvalue value problem for arbitrary $\epsilon>0$. We show in Section~\ref{spectrum}
that the linear stability operator, defined in an appropriate space of analytic functions, has a pure point
spectrum. In Section~\ref{PosEig} it is proven that there are no discrete eigenvalues with non-negative real
part, except $\lambda=0$ that corresponds to the trivial translation mode. A set of discrete, purely
negative eigenvalues is calculated in Section~\ref{negative eigenvalues} as a function of $\epsilon$; they
smoothly extend the results found previously for $\epsilon=1$. The results suggests that as $\epsilon
\rightarrow 0$, the spectrum degenerates to the trivial translation mode and this limit is discussed in
detail in Section~\ref{eps0}. Section~\ref{EigenFunc} contains a discussion of the eigenfunctions belonging
to these eigenvalues, and Section~\ref{Concl} contains the conclusions. Some part of our analysis exploits
general results on the asymptotic behavior of the coefficients of Taylor expansions. These results are
presented in an appendix.


\section{Reformulation by conformal mapping} \label{reformulation}

In this section, we collect results and notations from
\cite{meul05,eber07} that will be used in later
sections.

\subsection{Problem formulation and rescaling} \label{scaling}

We consider a compact ionized domain $\cal D$ in the $(x, y)$-plane. We assume that the net charge on the
domain vanishes (i.e., it contains the same number of electrons and positive ions). The domain moves in an
external field that far from the domain asymptotically approaches
\begin{equation} \label{Ametrin}
{\bf E}_\infty = -|{\bf E}_\infty | ~\hat {\bf x} \, .
\end{equation}
Here $\hat {\bf x}$ is the unit vector in $x$-direction, and $|{\bf E}_\infty |$ sets the scale of ${\bf E}$
and thus of the potential $\varphi$. As length scale we take
\begin{equation} \label{Antimonit}
R = \sqrt {\frac {| \cal D |}{\pi} } \, .
\end{equation}
where $| \cal D |$ is the area of $\cal D$, which is known to be conserved. This follows from the charge
neutrality of the streamer since
\begin{equation}\label{area1}
0=\int_{\cal D} dx~dy~\nabla\cdot{\bf E}=\int_{\partial\cal D} ds~{\bf n}(s)\cdot{\bf E}
=-\int_{\partial{\cal D}}ds~v_n,
\end{equation}
where in the last step we inserted Eq.~(\ref{Amazonit}) for the normal velocity of the boundary. Since
\begin{equation}\label{area2}
\int_{\partial{\cal D}}ds~v_n = \partial_t |{\cal D}|,
\end{equation}
the area is conserved, irrespective of the precise charge distribution in the interior\footnote{We remark
that the argument is straight forward to generalize to three spatial dimensions. Therefore the volume of a
charge neutral object with surface velocity ${\bf v}\propto {\bf E}^+$ in three spatial dimensions is
conserved as well.}. Also introducing the time scale $R/ | E_\infty |$, we rescale the basic equations to
the dimensionless form
\begin{eqnarray}
\label{Apatit}
\Delta \varphi & = & 0 \,, \quad (x, y) \notin \cal D \\
\label{Aquamarin}
v_n & = & {\bf n} \cdot (\nabla \varphi)^+ \\
\label{Aragonit} \varphi^+ & = & \epsilon \, {\bf n} \cdot (\nabla \varphi)^+ \,.
\end{eqnarray}
The only remaining parameter in the rescaled problem is
\begin{equation} \label{Girasol}
\epsilon=\ell/R.
\end{equation}
The boundary condition at infinity after rescaling takes the form
\begin{equation} \label{Azurit}
\varphi \to x + {\rm const} \quad {\rm{for}} \quad \sqrt {x^2 + y^2} \to \infty \, .
\end{equation}

\subsection{Conformal mapping} \label{conformal}

We now identify the physical $(x, y)$-plane with the closed complex plane $z = x + i y$, and we introduce a
conformal map $f (\omega, t)$ that maps the unit disk ${\cal U}_\omega$ in the $\omega$-plane to the
complement of $\cal D$ in the $z$-plane, with $\omega = 0$ being mapped on $z = \infty$,
\begin{equation} \label{Bergkristall}
z = f (\omega, t) = \frac {a_{-1} (t)}{\omega} + \hat f (\omega, t), \quad a_{-1} (t) > 0\, .
\end{equation}

We further define a complex potential $\Phi (\omega, t)$ obeying
\begin{equation} \label{Bernstein}
{\rm Re}[\Phi(\omega, t)] = \varphi(f(\omega, t)) ~~~\mbox{for }\omega \in {\cal U}_\omega \, .
\end{equation}
The boundary condition (\ref{Azurit}) and the Laplace equation (\ref{Apatit}) enforce the form
\begin{equation} \label{Beryll}
\Phi (\omega, t) = \frac {a_{-1} (t)}{\omega} + \hat \Phi (\omega, t)
\end{equation}
with $\hat \Phi$ being holomorphic for $\omega \in {\cal U}_\omega$.

The two boundary conditions (\ref{Aquamarin}), (\ref{Aragonit}) take the form
\begin{eqnarray}
\label{Biotit} {\rm Re} \left[ \frac {\partial_t f}{\omega \partial_\omega f} \right] &=& {\rm Re}
\left[\frac {\omega\partial_\omega \Phi}{| \partial_\omega f |^2}\right] ~~~\mbox{for } \omega \in \partial
\, {\cal U}_\omega \, ,
\\
\label{Blauquarz} | {\partial_\omega f} | ~{\rm Re}
[\Phi] &=& -\epsilon \,{\rm Re} [\omega
\partial_\omega \Phi] ~~~\mbox{for } \omega \in \partial \, {\cal U}_\omega \, ,
\end{eqnarray}
which completes the reformulation of the moving boundary problem (\ref{Apatit})--(\ref{Azurit}) by conformal
mapping.

We will restrict the analysis here to initial conditions $\hat f (\omega, 0)$ holomorphic in some domain
${\cal U}^\prime_0 \supset {\cal U}_\omega$. In part II~\cite{partII} of this paper sequence, we will give
evidence that analyticity on $\overline{\cal U}_\omega$ is preserved in time, though the distance of the
domain of analyticity ${\cal U}'_t$ to $\partial {\cal U}_\omega$ shrinks with time. The streamer boundary
$\partial {\cal D}$, which is the image of boundary $\partial \mathcal{U}_\omega$ under $f(\omega, t)$, will
turn out to be analytic and therefore smooth.
Similar analytic representations exist for the entire
class of 2-D Laplacian growth,
with details depending on the type of boundary condition, geometry and
asymptotic conditions at infinity. For the classic viscous fingering problem,
Polubarinova-Kochina
\cite{Polubarinova} and Galin \cite{Galin}
use a representation that coincides with the one given above in
the unregularized case $\epsilon=0$.

\subsection{Linear perturbation of moving circles} \label {linear}

It is easily seen that equations (\ref{Biotit}), (\ref{Blauquarz}) allow for the simple solution
\begin{eqnarray} \label{Calcit}
\begin{array}{rcrcr}
f^{(0)} (\omega, t) & = & \displaystyle \frac {1}{\omega} + \frac {2 t}{1 + \epsilon},
\\
\Phi^{(0)} (\omega, t) & = & \displaystyle \frac {1}{\omega} - \frac {1 - \epsilon}{1 + \epsilon}\  \omega,
\end{array}
\end{eqnarray}
which in physical space describes circles of radius $1$ moving with constant velocity $2 / (1 + \epsilon)$
in $x$ direction. (We recall that the radius was scaled to unity in Section~\ref{scaling}.) We note that
relaxing the analyticity conditions on $f (\omega, t)$ on $|\omega|=1$, one can obtain another set of
uniformly translating solutions, as recently discovered \cite{prok08}. The present paper is restricted to
perturbations of the steady circle that retain the imposed analyticity of the streamer shapes, and hence
analyticity of $f$ (as well as $\Phi$) on $|\omega|=1$.

As the area is conserved (as shown in subsection~\ref{scaling}), the residue $a_{-1} = 1$ does not change to
linear order in the perturbation. We therefore can use the ansatz
\begin{eqnarray} \label{Charoit}
\begin{array}{rcrcr}
f(\omega, t) & = & f^{(0)} (\omega, t) + \eta~ \beta(\omega, t) \, ,
\\[1.5ex]
\Phi(\omega, t) & = & \Phi^{(0)} (\omega, t) + \eta ~ \frac 2{1+\epsilon} \chi(\omega, t) \, ,
\end{array}
\end{eqnarray}
where $\eta$ is a small parameter, and $\beta (\omega, t)$, $\chi (\omega, t)$
are holomorphic in ${\cal
U}_{\omega}$.
A first order expansion of Eqs.~(\ref{Biotit}), (\ref{Blauquarz}) in $\eta$ yields the
following boundary conditions for the analytic functions $\beta (\omega, t)$ and $\chi (\omega, t)$ on
$|\omega|=1$:
\begin{eqnarray}
\label{Nigrin}
\begin{array}{rcrcr}
Re[\omega \partial_{\tau} \beta- \omega \partial_{\omega} \beta] & =
& Re[-\omega\partial_{\omega} \chi] \, ,
\\[1.5ex]
\frac {\epsilon}{2} Re \left [ (\omega+\frac{1}{\omega}) \omega^{2}
\partial_{\omega}\beta \right ]  & = & Re[\epsilon\omega\partial_{\omega} \chi +\chi]\, ,
\end{array}
\end{eqnarray}
where we rescaled time as
\begin{equation} \label{Dioptas}
\tau = \frac 2{1+\epsilon} ~t.
\end{equation}
Since the left and right sides of each of the two equations in (\ref{Nigrin}) are real parts of analytic
functions and each is assumed {\it a priori} continuous upto the boundary, they can differ everywhere in
$\omega$ by at most an imaginary constant. Evaluation at $\omega=0$ shows this constant to be zero for the
first of the two equations. Elimination of $\chi$ results in the linear PDE:
\begin{equation} \label{Citrin}
{\cal L}_\epsilon \, \beta = 0
\end{equation}
with the operator
\begin{equation} \label{Diamant}
{\cal L}_\epsilon = \frac {\epsilon}{2}\  \partial_\omega \  (\omega^2 - 1)\  \omega \, \partial_\omega +
\epsilon \, \partial_\omega \,\omega \,  \partial_\tau + \partial_\tau - \partial_\omega \,.
\end {equation}

We note that ${\cal L}_\epsilon$ is of similar structure as the operator resulting from a linear
stability analysis of translating circles in the context of void electromigration
\cite{mah-96,Cummingsetal}. The main difference here is the occurrence of the mixed derivative
$\partial_\omega \,\omega \,  \partial_\tau$.

\subsection{Formulation of the eigenvalue problem} \label{precise formulation}

To motivate our formulation of the eigenvalue problem, we note some results on the temporal evolution of
infinitesimal perturbations. In \cite{eber07}, the equation ${\cal L}_\epsilon \, \beta = 0$ was solved as
an initial value problem for the special value $\epsilon = 1$. It was found that any initial perturbation
$\beta (\omega, 0)$ holomorphic in ${\cal U}' \supset
{\cal U}_\omega$ for $\tau \to \infty$ is
exponentially convergent to some constant. This results from the expansion
\begin{equation} \label{Disthen}
\beta(\omega, \tau) = \sum_{n = 0}^{\infty} g_n \, \beta_{\lambda_n}^{(1)}(\omega)\; e^{\lambda_n \tau},
\end{equation}
with
\begin{equation}
\label{Dolomit} \lambda_n  = - n, \quad n \in \mathbb{N}_0, ~~~\mbox{for }\epsilon=1.
\end{equation}
The coefficients $g_n$ and the eigenfunctions\footnote{Note that the exponent $-\lambda_n$ in (\ref{eps1})
is correct while $+\lambda_n$ in Eq.~(4.20) in \cite{eber07} is a typo.}
\begin{equation}
\beta_{\lambda_n}^{(1)}(\omega)= \int_0^\omega
\frac{x~dx}{\omega^2}~\left(\frac{x-1}{x+1}\right)^{-\lambda_n}\label{eps1}
\end{equation}
are determined by an expansion of $(2+\omega\partial_\omega)\beta(\omega,0)$ in powers of
$(1-\omega)/(1+\omega)$. For $n>0$ the eigenfunctions (\ref{eps1}) are singular at $\omega=-1$, though
$\beta(\omega,\tau)$ is not. The expansion (\ref{Disthen}) is convergent in a domain ${\cal D}_\tau$
expanding in time that eventually includes every point in $\bar{\cal U}_\omega \setminus \{-1 \}$. For large
$\tau$ the region where the expansion is invalid, shrinks to $\omega=-1$ exponentially. This region is
measured by the new scale $\eta_1(\omega,\tau) =(1+\omega)e^\tau$, and the expansion (\ref{Disthen}) is
valid if $\eta_1$ is large. For $\eta_1 \le O(1)$ the perturbation for $\tau\to\infty$ behaves as $\beta
(\omega, \tau) \rightarrow F_0 (\eta_1) + O(e^{-\tau})$ where $F_0$ is some analytic function of its
argument, depending on $\beta (\omega, 0)$.

For an analytic initial condition on $\mathcal{U}'$, with a lone branch point singularity $\omega_s$ in
$|\omega| > 1$ not on the positive real axis, the emergence of this new scale near $\omega = -1$ can be
related to the approach of this complex singularity towards $-1$ exponentially in $\tau$ for large $\tau$.
Asymptotic arguments that will be presented in part II~\cite{partII} suggest that this behavior is generic
for all $\epsilon>0$. The analysis is based on the linear the evolution equations for $b_k$, where
$$\beta (\omega, \tau) =\sum_{k=0}^\infty b_k (\tau) \omega^k. $$
For $k \gg e^{\tau}$, we find the asymptotic relation
$$b_k \sim (-1)^k k^{-\alpha} h(\tau) \exp \left [-k f(\tau) \right ], $$
where
$$f(\tau) = \log \left [ \frac{1+C e^{-\tau}}{1-C e^{-\tau}} \right ],
~~{\rm with} ~C = \frac{\omega_s+1}{\omega_s-1}.$$ For $\omega_s \notin (1, \infty)$, $f(\tau)$ stays finite
and approaches 0 exponentially in $\tau$ for large $\tau$. If $\omega_s \in (1, \infty)$, $f(\tau)$
increases monotonically to $\infty$ for $\tau \in (0, \tau_c)$ where $e^{-\tau_c}=1/C$. For $\tau> \tau_c$,
$f(\tau)$ decreases monotonically and approaches 0 exponentially in $\tau$ as $\tau \rightarrow \infty$. In
either case, from the known relation between Taylor series coefficients and the location of the closest
singularity of an analytic function (see the appendix), it follows that $f(\tau) \sim e^{-\tau}$ as $\tau
\rightarrow \infty$ implies that $\beta $ has a singularity approaching $\omega = -1$ exponentially in
$\tau$ for large $\tau$. This feature is retained for any other isolated initial singularities as well,
though $k^{-\alpha}$ is replaced by a more complicated dependence in $k$. Since the problem is linear, the
evolution of a distribution of initial singularities can be understood from the linear superposition
principle.

This suggests that for any $\epsilon>0$, as for $\epsilon=1$, $\beta (\omega, \tau)$ has a collapsing scale
$(1+\omega) e^{\tau} $, and an expansion of the type (\ref{Disthen}) cannot be valid in this neighborhood of
$\omega=-1$.

Thus, in seeking an eigenfunction by substituting
\begin{equation} \label{Epidot}
\beta(\omega, \tau) = \beta_\lambda^{(\epsilon)}(\omega) ~e^{\lambda \tau} \,,
\end{equation}
into (\ref{Citrin}), (\ref{Diamant}), it is appropriate to allow $\beta^{(\epsilon)}_{\lambda}$ to be
singular at $\omega=-1$. Indeed, substituting the form (\ref{Epidot}) reduces Eqs.~(\ref{Citrin}),
(\ref{Diamant}) to the eigenvalue problem
\begin{equation}
\label{Falkenauge}
L(\epsilon,\lambda)~ ~\beta_\lambda^{(\epsilon)} (\omega) = 0 , \end{equation}
\begin{equation}
\label{Zirkonia} L(\epsilon,\lambda)= \frac{\epsilon\,(\omega^2-1) \omega }2 \,\partial^2_\omega +
\left(\frac{\epsilon \,(3\omega^2-1)}2
-1\right)\partial_\omega+\lambda(1+\epsilon+\epsilon\omega\partial_\omega) .
\end{equation}
Evidently this ODE has three regular singular points, namely $\omega =0$ and $\omega = \pm 1$. The
independent solutions at these points for $\epsilon>0$ are in leading order
\begin{eqnarray} \label{Fluorit}
\beta_\lambda^{(\epsilon)}(\omega) & \sim  \displaystyle\left\{
\begin{array}{ll}
{\displaystyle \omega^0 \atop  \displaystyle\omega^{-2/\epsilon}} &~~~\mbox{for }\omega \to 0,
\end{array}  \right.
\\
\label{Granat} \beta_\lambda^{(\epsilon)}(\omega) & \sim  \displaystyle \left \{
\begin{array}{ll}
{\displaystyle (1 \mp \omega)^0 \atop \displaystyle (1 \mp \omega)^{1/\epsilon\,\mp \lambda}} &~~~\mbox{for
}\omega \to \pm 1,
\end{array} \right.
\end{eqnarray}
We require the eigenfunctions $\beta_{\lambda}^\epsilon$ to be solutions of (\ref{Falkenauge}) that are
analytic in $\omega =0$ and $\omega =1$. This is also the natural choice from a physical point of view since
it is the right half of the circle, ${\rm Re}[\omega] > 0$, that corresponds to the physically interesting
tip of the streamer. In general, eigenfunctions cannot be expected to be regular at all three points.
Starting with a function regular at $\omega=0$, we cannot generally require regularity at both points
$\omega=\pm1$ by adjusting the single parameter $\lambda$. As shown in subsection~\ref{pureim}, the only
eigenfunction regular at all three points is the trivial translation mode
\begin{equation} \label{Heliotrop}
\lambda_0 = 0, \qquad \beta_0^{(\epsilon)} (\omega) = {\rm const.}
\end{equation}

As noted above, an operator similar to ${\cal L}_\epsilon$ (\ref{Diamant}) occurs in the problem of void
electromigration, see section 4.1.3 in \cite{Cummingsetal}. Again an eigenvalue analysis would yield a
second order linear operator with three singular points at $\omega=0$ and $\pm{1}$ and therefore the
eigenmodes in general cannot be regular at all three singular points. It is interesting to note that the
authors \cite{Cummingsetal} conclude that their problem is unstable because the initial value problem for
large time is singular at $\omega=-1$. In the current problem, the solution~\cite{partII} of the initial
value problem is not singular at $\omega=-1$; the singularity of the eigenfunctions does not reflect the
true behavior of solution since, as has been pointed out earlier, there is an anomalous contracting scale
$e^{\tau} (1+\omega)$ near the back of the bubble. Whether or not there is an analogous contracting scale
for the void electromigration problem \cite{Cummingsetal} remains an interesting question. This anomalous
scale shows up when the limiting processes $\lim_{\omega \rightarrow -1}$ and $\lim_{\tau \rightarrow
+\infty}$ do not commute for the solution of the initial value problem.


\section{Discreteness of the spectrum}\label{spectrum}

We define $\lambda$ to be in the spectrum, if the linear operator $L (\epsilon, \lambda)$ does not have a
bounded inverse in the class of functions $f$ that are analytic in an arbitrary compact connected set
$\mathcal{V} \subset \mathcal{U}^\prime \setminus \{ -1 \} $ that contains the whole line $[0,1]$ in its
interior. $\lambda$ is in the discrete spectrum if (\ref{Falkenauge}) has a nonzero solution
$\beta_{\lambda}^{(\epsilon)} (\omega) $ that is analytic in any such domain $\mathcal{V}$. We now argue
that if $\lambda$ is not in the discrete spectrum, then $L (\epsilon, \lambda)$ has a bounded inverse, {\it
i.e.} there is only a discrete spectrum in this problem.

To determine $L^{-1} $, we solve the equation
\begin{equation} \label{3-33}
L (\epsilon, \lambda) g = h
\end{equation}
for a given $h$ analytic in $\mathcal{V}$, imposing the condition that also $g$ is analytic in
$\mathcal{V}$. The solutions of the homogeneous equation $L(\epsilon,\lambda)f=0$ that are regular at
$\omega=0$ or $\omega=1$ will be denoted by $f_1(\omega)$ or $f_2(\omega)$, respectively. It follows from
Eqs.~(\ref{Fluorit}), (\ref{Granat}) that these functions are determined uniquely up to a multiplicative
constant. In the exceptional case where both independent solutions are regular at $\omega=1$, $\lambda$
belongs to the discrete spectrum, see Section~\ref{negative eigenvalues}. A standard calculation shows that
Eq.~(\ref{3-33}) is solved by
\begin{equation}\label{3-34}
g(\omega)=\frac1{C(\lambda,\epsilon)}\int_0^\omega d\omega'~G(\omega,\omega')~h(\omega')+a_1[h]~f_1(\omega),
\end{equation}
where
\begin{equation}\label{3-35}
G(\omega,\omega')=\frac{\omega'^{2/\epsilon}}{(1-\omega')^{1/\epsilon-\lambda}(1+\omega')^{1/\epsilon+\lambda}}
~\left[f_2(\omega)f_1(\omega')-f_1(\omega)f_2(\omega')\right],
\end{equation}
and the coefficient $a_1[h]$ is a functional of $h(\omega')$. $C(\lambda,\epsilon)$ does not vanish since
otherwise the Wronskian $f_1~\partial_\omega f_2-f_2~\partial_\omega f_1$ vanishes identically and $\lambda$
is part of the discrete spectrum. It is easily seen that Eqs.~(\ref{3-34}), (\ref{3-35}) render $g(\omega)$
analytic in $\omega=0$, and this condition eliminates
any contribution of the form $a_2 [h]~f_2(\omega)$.

Analyticity at $\omega=1$ is enforced by a proper choice of $a_1 [h]$. To make the analysis explicit, in
addition to $f_2(\omega)$, we introduce another solution to $L[\epsilon, \lambda] f=0$ by requiring
\begin{equation}\label{3-36}
f_3(\omega)=(1-\omega)^{1/\epsilon-\lambda}~\hat f_3(\omega),
\end{equation}
where $\hat f_3(\omega)$ is analytic at $\omega=1$. Using this form of $f_3(\omega)$, we exclude the case
$\frac1\epsilon-\lambda\in\mathbb{Z}^+$, that will be discussed later. Writing $f_1(\omega)$ as
\begin{equation}\label{3-37}
f_1(\omega)=c_2 f_2(\omega)+c_3 f_3(\omega),
\end{equation}
we find that $G(\omega,\omega')$ from Eq.~(\ref{3-35}) takes the form
\begin{equation*}
G(\omega,\omega')=\frac{c_3 \omega'^{\;2/\epsilon}}{(1+\omega')^{1/\epsilon+\lambda}}~ \left[f_2(\omega)\hat
f_3(\omega')-\left(\frac{1-\omega'}{1-\omega}\right)^{\lambda-1/\epsilon} \hat
f_3(\omega)f_2(\omega')\right].
\end{equation*}
Evidently the first part in the square brackets for $\omega\to1$ yields a regular contribution to
$g(\omega)$ from Eq.~(\ref{3-34}). The contribution to $\int G\;h$ that is singular in $\omega=1$ has the
form
\begin{equation*}
-c_3 f_3(\omega) \int_0^\omega d\omega'~(1-\omega')^{\lambda-1/\epsilon} H(\omega'),
\end{equation*}
where
\begin{equation}\label{3-38}
H(\omega')=\frac{\omega'^{\;2/\epsilon}}{(1+\omega')^{1/\epsilon+\lambda}}~f_2(\omega')~h(\omega')
\end{equation}
is regular at $\omega'=1$. If Re\;$\lambda-\frac1\epsilon>-1$, we can write
\begin{eqnarray}\label{3-39}
\lefteqn{-c_3 f_3(\omega) \int_0^\omega d\omega'~(1-\omega')^{\lambda-1/\epsilon} H(\omega')} \nonumber\\
&=& -c_3 f_3(\omega) \int_0^1 d\omega'~(1-\omega')^{\lambda-1/\epsilon} H(\omega') \nonumber\\
&& +c_3 \hat f_3(\omega) \int_\omega^1 d\omega'~\left(\frac{1-\omega'}{1-\omega}\right)^{\lambda-1/\epsilon}
H(\omega').
\end{eqnarray}
The second part is regular at $\omega=1$ and the singular first part is canceled by the choice
\begin{equation}\label{3-40}
a_1[h] = \int_0^1 d\omega'~(1-\omega')^{\lambda-1/\epsilon} H(\omega').
\end{equation}
We note that this result is valid also for $\lambda=\frac1\epsilon+n$, $n\in\mathbb{N}$, where $f_3(\omega)$
instead of being of the form (\ref{3-36}) shows a logarithmic singularity.

If $-n>\;{\rm Re}\;\lambda-\frac1\epsilon>-n-1$, $n\in\mathbb{N}$, we carry through $n$ subtractions of
$H(\omega')$ at $\omega'=1$, defining
\begin{equation}\label{3-41}
\left[H(\omega')\right]_n=H(\omega')-\sum_{j=0}^{n-1} H_j~(1-\omega')^j,
\end{equation}
so that $\left[H(\omega')\right]_n\sim {\rm const}~(1-\omega')^n$. A short calculation shows that the
singular part of $\int G~h$ is canceled by the choice
\begin{equation}\label{3-42}
a_1[h]=\int_0^1 d\omega'~(1-\omega')^{\lambda-1/\epsilon} \left[H(\omega')\right]_n +\sum_{j=0}^{n-1}
\frac{H_j}{\lambda-\frac1\epsilon+j+1}.
\end{equation}
The expressions above clearly remain valid when $\frac{1}{\epsilon} - {\rm Re}\; \lambda = n $, except when
$\frac{1}{\epsilon} - \lambda = n$, a positive integer.

When $\frac1\epsilon-\lambda = n $ is a positive integer, from well-known theory \cite{WW} for regular
singular points, instead of (\ref{Granat}), the solutions $f_1$ and $f_2$ as defined earlier must have the
following local representation near $\omega=1$:
\begin{equation}
\label{f1exp}
f_1(\omega) = C_1 (1-\omega)^n B_1 (\omega) \log (1-\omega) + B_2 (\omega),
\end{equation}
\begin{equation}
\label{f2exp}
f_2 (\omega) = (1-\omega)^n B_1 (\omega),
\end{equation}
where $B_1$ and $B_2$ are analytic at $\omega=1$. If $f_1$ and $f_2$ are independent, as they are when
$\lambda$ is not in the discrete spectrum, then $C_1 (\epsilon, \lambda) \ne 0$. 

We now define
\begin{equation}
\label{Halt}
H(\omega) = B_2 (\omega) (1+\omega)^{n-2/\epsilon} \omega^{2/\epsilon}
h(\omega),
\end{equation}
while $H_j$ is still defined by the expression (\ref{3-41}). It is also convenient to define
\begin{equation}
\label{Galt} Q(\omega) = B_1 (\omega) (1+\omega)^{n-2/\epsilon} \omega^{2/\epsilon} h(\omega),
\end{equation}
Note that each of $H$ and $Q$ are analytic at $\omega=1$. Straight forward calculation based on (\ref{3-34})
shows that the possibly singular part of $g(\omega)$ at $\omega=1$ is given by
\begin{multline*}
-\frac{H_{n-1} }{C} \ln (1-\omega) B_1 (\omega) (1-\omega)^n \\
+ f_1 (\omega) \left ( a_1 - \int_0^1
d\omega'\; \frac{{\omega'}^{2/\epsilon} f_2 (\omega') h(\omega')}{
C(\epsilon, \lambda) (1-\omega')^n (1+\omega')^{2/\epsilon -n} } \right ) \\
+ \frac{C_1 f_2 (\omega)}{C} \int_1^\omega d\omega'~Q(\omega') \ln \frac{1-\omega'}{1-\omega}
\end{multline*}
The last term is analytic at $\omega=1$. The singularity vanishes if we choose
\begin{equation}
\label{a1choice} a_1 = \frac{H_{n-1} }{C_1 C } + \int_0^1 d\omega'~ \frac{{\omega'}^{2/\epsilon} f_2
(\omega') h(\omega')}{ C(\epsilon, \lambda) (1-\omega')^n (1+\omega')^{2/\epsilon -n} }  .
\end{equation}

For any $\lambda$ for which $C(\lambda, \epsilon) \ne 0$, using the explicit expression (\ref{3-34}) with
$a_1$ determined from (\ref{3-40}), (\ref{3-42}) or (\ref{a1choice}), whatever the case may be, we have in
the domain $\mathcal{V}$,
\begin{equation}
\label{Linvbound}
\| g \|_{\infty} \le C \| h \|_{\infty},
\end{equation}
This conclusion follows from observing the properties of the integrand and noting that the $H_j$, $j=1,...n$
are bounded by some multiples of $\sup_{\omega \in \mathcal{V}}  |h (\omega)|$, since they involve only a
finite number of derivatives of $h$ at $\omega =1$. Since the boundary $\partial \mathcal{V} $ is at a
finite distance from $\omega=1$, the derivatives $\partial_\omega^j h\big|_{\omega=1}$ by Cauchy's theorem
are bounded by $b_j\;\sup_{\omega \in \mathcal{V}} |h (\omega)|$
where $b_j$ is independent of $h$.
Hence, we have shown\footnote{Note from definition, $\lambda$ is not in the
spectrum
if the resolvent $ L^{-1}$ is bounded.}
that $\lambda$ is in the spectrum only if $C(\lambda, \epsilon)=0$, {\it i.e.}
we can only have discrete spectrum in this problem.


\section{Absence of eigenvalues with positive real part and of purely imaginary eigenvalues} \label{PosEig}

\subsection{Purely positive eigenvalues}

Real eigenvalues $\lambda > 0$ easily are excluded. Substituting into Eq.~(\ref{Falkenauge}) the power
series
\begin{equation} \label{Holzstein}
\beta_\lambda^{(\epsilon)}(\omega) = \sum_{k=0}^{\infty} b_k \, \omega^k \, ,
\end{equation}
which converges for $|\omega| < 1$ due to the location of the regular singular points, we find the recursion
relation
\begin{equation}
\label{Howlith} b_k = 2 \lambda \, \frac {1 + \epsilon k}{k (2 + \epsilon k)} \, b_{k -1}\ + \epsilon
\,\frac {k - 2}{2 + \epsilon k} \, b_{k - 2}~~~\mbox{for } k \geq {2} \, ,
\end{equation}
and
\begin{equation} \label{Zinnstein}
b_{1}  =  2\lambda\;\frac{1+\epsilon}{2+\epsilon} b_{0} \, .
\end{equation}
We choose $b_{0} = 1$ as initial value.

For $\lambda > 0$, evidently all $b_k$ are positive, and $b_{k}$ obeys the bound
\begin{displaymath}
b_{k} > \epsilon \frac {k - 2}{2 + \epsilon k} ~b_{k - 2}\, ,
\end{displaymath}
and therefore
\begin{equation}\label{33}
b_{k} > \frac {\Gamma \left(\frac k 2\right)}{\Gamma\left(1+\frac 1 \epsilon+\frac k 2\right)}~{\rm
const.}>0.
\end{equation}
For $k \gg {1}/{\epsilon}$, this yields the lower bound
\begin{displaymath}
b_{k} > {\rm const} ~ k^{-1 - 1/\epsilon} \, ,
\end{displaymath}
which shows that a sufficiently high derivative of $\beta_\lambda^{(\epsilon)} (\omega)$ (\ref{Holzstein})
diverges for $\omega = 1$, which contradicts the regularity requirement.

\subsection{Eigenvalues with positive real part}

To eliminate eigenvalues $\lambda=\mu+i\nu$ with $\mu>0$ needs more refined arguments. We first derive an
inequality replacing (\ref{33}) above. Motivated by (\ref{33}),
we rewrite the recursion relation
(\ref{Howlith}) in terms of
\begin{equation}
\label{34}
c_k=\frac{\Gamma\left(1+\frac 1 \epsilon+\frac k 2\right)}{\Gamma \left(\frac k 2\right)}~b_k,~~~k\ge1.
\end{equation}
This yields
\begin{equation}\label{35}
c_k=\lambda\;g_k\;c_{k-1}+c_{k-2},~~~k\ge3,
\end{equation}
where
\begin{equation}\label{36}
g_k\equiv\frac{2(1+\epsilon k)\;\Gamma\left(\frac{k-1}2\right)\;\Gamma\left(1+\frac 1 \epsilon+\frac k
2\right)}{k(2+\epsilon k)\;\Gamma\left(\frac k 2\right)\;\Gamma\left(\frac 1 2 +\frac 1 \epsilon+\frac k
2\right)}=\frac 2 k \;\left(1+{\cal O}\left(\frac 1 k\right)\right).
\end{equation}
We now multiply (\ref{35}) by $c^*_{k-1}$ and take the real part. With the notation
\begin{equation}\label{37}
r_k={\rm Re}\left\{c_k c^*_{k-1}\right\},
\end{equation}
we get the relation
\begin{equation}\label{38}
r_k=\mu\;g_k\;|c_{k-1}|^2+r_{k-1},~~~k\ge3.
\end{equation}
Since $\mu>0$ and
\begin{displaymath}
r_2={\rm Re}\left\{c_2 c^*_1\right\}=2\mu|\lambda|^2~\frac{(1+\epsilon)(1+2\epsilon)}{(2+\epsilon)^2}~
\frac{\Gamma\left(\frac32+\frac 1 \epsilon\right)~\Gamma\left(2+\frac 1 \epsilon\right)}
{\Gamma\left(\frac12\right)}>0,
\end{displaymath}
the $r_k$ form an increasing series of positive numbers bounded by
\begin{equation}\label{39}
r_k\ge r_2>0.
\end{equation}
The recursion relation (\ref{38}) formally is solved as
\begin{equation}\label{40}
r_k=\mu\sum_{j=3}^k g_j\;|c_{j-1}|^2+r_2,~~~ k\ge3.
\end{equation}
Using now the relation
\begin{displaymath}
|c_k|^2+|c_{k-1}|^2=2r_k+|c_k-c_{k-1}|^2\ge2r_k,
\end{displaymath}
we find the bound
\begin{equation}\label{41}
\frac{|c_k|^2+|c_{k-1}|^2}2\ge\mu\sum_{j=3}^k g_k\;|c_{j-1}|^2+r_2.
\end{equation}
We note that this bound is positive and increases monotonically.

We now recall that by definition of the eigenfunctions, the only singularity in the complex plane is of the
form
\begin{displaymath}
(1+\omega)^{1/\epsilon+\lambda}\;f(\omega),
\end{displaymath}
where $f(\omega)$ is regular in $|\omega|\le 1$, including $\omega=-1$. A standard result on the relation of
power series coefficients to the closest complex singularity (see the appendix) is, that the asymptotic
behavior of $b_k$ of the expansion (\ref{Holzstein}) satisfies
\begin{displaymath}
|b_k|\sim B_\infty~k^{-\mu-1/\epsilon-1},
\end{displaymath}
where $B_\infty$ is some constant. In view of (\ref{34}), this implies
\begin{displaymath}
|c_k|\sim {\rm const}~ 2^{-1-1/\epsilon}~k^{-\mu} \to 0
\end{displaymath}
for $k\to\infty$ which contradicts the bound (\ref{41}). We thus conclude that there are no eigenvalues
$\lambda$ with Re~$\lambda>0$.

\subsection{Purely imaginary eigenvalues}\label{pureim}

Now consider the possibility of a purely imaginary eigenvalue $\lambda=i\nu$ with $\nu $ real. From the
recursion relation, it is clear that $\nu=0$ corresponds to the translation mode. So we only consider the
case $\nu \ne 0$. From complex conjugation symmetry, it is clear that if $\lambda = i \nu$ is an eigenvalue,
so is $\lambda = - i \nu$. Therefore, we may assume without any loss of generality that $\nu > 0$.

We introduce
\begin{equation}
{\hat b}_k =i^{-k}b_k.
\end{equation}
The recursion relation (\ref{Howlith}) takes the form
\begin{equation}\label{RecIm}
\hat b_k = 2 \nu \, \frac {1 + \epsilon k}{k (2 + \epsilon k)} \, \hat b_{k -1}\ - \epsilon \,\frac {k -
2}{2 + \epsilon k} \, \hat b_{k - 2},~~~\hat b_0=1,~\hat b_{-1}=0 ,
\end{equation}
which shows that $\hat b_k$ is real for any $k$. Thus the function
\begin{equation}
\hat\beta(\hat\omega)=\sum_{k=1}^\infty \hat b_k
\hat\omega^k=\beta_\lambda^{(\epsilon)}(\omega),~~~\hat\omega=i\omega,
\end{equation}
is real for real $\hat\omega$, and the reflection principle guarantees
$$ \hat\beta^*(\hat\omega^*)=\hat\beta(\hat\omega).$$
Thus $\hat\beta(\hat\omega)$ either is singular both at $\hat\omega=\pm i$, corresponding to $\omega=\mp1$,
or is entire. When $\lambda = i \nu$ is some eigenvalue, we cannot have a singularity at $\hat\omega = -{i}$
and so $\hat\beta(\hat\omega)$ must be entire. We will now show that this is impossible.

First, note from the recursion relation that $\hat\beta(\hat\omega)$ cannot be a polynomial since if $b_{k}
= 0 = b_{k-1}$, then so must $b_{k-2}$ and all the previous coefficients. Choose $k_0$ so large that for $k
\ge k_0 \ge 4$
\begin{equation}
\label{k0constr}
\frac{\epsilon k}{1+\epsilon (k+2)} \ge \frac{1}{2} ~~~\\, ~~~
\frac{2+\epsilon (k+2)}{\epsilon k} \le 2
\end{equation}
We choose a specific $\rho $ large enough so that
\begin{equation}
\label{rhobb}
\rho \ge  4 ~~~,~~~
\frac{\rho}{2 \nu}  > 4
\end{equation}

Since $\sum_{k=k_0}^\infty {\hat b}_k {\hat \omega}^k$
is an entire function, it follows that for any $k_0$,
there exists a constant $M > 0$ so that
\begin{equation}\label{boundbarBk}
\rho^k |\hat b_k| \le M~~~{\rm for}~~k \ge k_0
\end{equation}
We redefine $M$ in the relation (\ref{boundbarBk})
to be the least upper-bound for $\rho^k |{\hat b}_k|$.
Note that since $\hat\beta(\hat\omega)$ is not a polynomial,
$M$ cannot be zero. We now introduce
\begin{equation}
d_k=\frac{\rho^k\,\hat b_k} M
\end{equation}
and rewrite recursion relation for $k \ge k_0+2$ as
\begin{equation}\label{RecIm2}
d_k = 2 \nu \, \frac {1 + \epsilon k}{k (2 + \epsilon k)} \,\rho\, d_{k -1}\ - \epsilon \,\frac {k - 2}{2 +
\epsilon k} \,\rho^2 \,d_{k - 2}.
\end{equation}
The bound (\ref{boundbarBk}) translates into
\begin{equation}\label{048}
|d_k|\le 1 ~~{\rm for} ~~k \ge k_0.
\end{equation}
Solving (\ref{RecIm2}) for $d_{k-1}$, and shifting the indices $k \rightarrow
k+2$, we obtain for $k \ge k_0$
\begin{equation}\label{049}
d_{k+1}= \frac{\rho (k+2)}{2\nu} \frac{\epsilon k}{1+\epsilon (k+2)} \left [ d_k + \left ( \frac{2+\epsilon
(k+2)}{\epsilon k \rho^2} d_{k+2} \right ) \right ]
\end{equation}
Since $M$ was the least upper bound for $\rho^k {\hat b}_k $, it follows that there exists some $k_* \ge
k_0$ so that
\begin{equation}\label{049.1}
|d_{k_*} | \ge \frac{1}{2}
\end{equation}
On using
(\ref{k0constr}), (\ref{rhobb}) and $| d_{k_*+2} | \le 1$,
it follows from (\ref{049}) that
\begin{equation}
|d_{k_*+1} | \ge (k_*+2) \left [ 1 - \frac{1}{4} \right ] > 1
\end{equation}
which is inconsistent with (\ref{048}).

We note  that this argument not only excludes imaginary eigenvalues, but it also shows that except for the
trivial translation mode $\beta_0^{(\epsilon)}(\omega)\equiv1$ corresponding to $\lambda =0$, there are no
eigenfunctions regular at all three points $\omega=0,\,\pm1$. This follows from simply replacing $\nu$ by
$-i\lambda$ (or $|\lambda|$, respectively) in the above analysis without necessarily restricting $\lambda$
to be imaginary.


\section{Calculation of negative eigenvalues for $\epsilon>0$} \label{negative eigenvalues}

We now concentrate on the infinite discrete set of negative eigenvalues $\lambda_{n} (\epsilon)$ which
continue the eigenvalues $\lambda_n (1)=-n$ found in \cite{meul05,eber07}; here the general case of
$\epsilon>0$ is considered while the limit $\epsilon\downarrow0$
is subject of section~\ref{eps0}.

Observing the parametric dependence of the operator $L$ in (\ref{Zirkonia}) on $\epsilon$ and $\lambda$, any
solution $\beta$ of the homogeneous equation $L(\epsilon, \lambda) \beta =0$ is analytic in $\lambda$ and
$\epsilon$, with the possible exception of $\epsilon=0$. Therefore the Wronskian of any two solutions is an
analytic function of $(\lambda, \epsilon)$, except at $\epsilon=0$. Since each eigenvalue $\lambda_n
(\epsilon) $ is determined as a zero of a particular Wronskian, it follows that it will change continuously
with $\epsilon$, except at $\epsilon=0$. Since eigenvalues $\left \{ \lambda_n (\epsilon) \right \}_n$ are
all real at $\epsilon=1$, as $\epsilon$ is decreased continuously from 1 towards 0, the only way eigenvalues
can become complex is through collision of erstwhile real eigenvalues, {\it i.e.}, through the existence of
a higher order zero of the Wronskian for some $\epsilon$. Such collisions are not observed in our numerical
calculation, consistent with the fact that all eigenspaces are one-dimensional, as is obvious from the
behavior of $\beta_\lambda^{(\epsilon)}(\omega)$ for $\omega\to0$ (\ref{Fluorit}). This suggests that
eigenvalues with negative real parts and  nonvanishing imaginary part are not possible, and therefore they
will not be considered in the ensuing.

For the present problem the calculation of $\lambda_n(\epsilon)$ as zeros of the Wronskian is feasible only
for $\epsilon$ not too small. Relying on the numerical solution of the ODE (\ref{Falkenauge}), with
decreasing $\epsilon$ this method rapidly breaks down since for $|\lambda| \ll 1/\epsilon$, the second
independent solution near $\omega=1$ shows only a very weak singularity, cf.~Eq.~(\ref{Granat}). It
therefore needs extreme numerical precision to determine reliably the solution that is regular at
$\omega=1$.

To circumvent this problem, we note that in the parameter space spanned by $(\epsilon,\lambda)$, there exist
special points where both independent solutions of Eq.~(\ref{Granat}) are regular at $\omega=1$. These
points are found on curves
%
\begin{equation} \label{Jade}
\lambda = \frac {1}{\epsilon} - m, \quad \frac {1}{\epsilon} < m \ \in \mathbb{N} \, ,
\end{equation}
where the general solution near $\omega=1$ can be written as \cite{WW}
\begin{equation} \label{Jaspis}
\beta_\lambda^{(\epsilon)}(\omega) = c_1 (\omega-1)^m B_1(\omega) + c_2 \Big(A(\epsilon) \,
(1-\omega)^{m}\,B_1(\omega)\; \ln (\omega-1)+ B_2(\omega)\Big)\, .
\end{equation}
Here $c_{1,2}$ are arbitrary constants, $B_{1,2}(1)=1$ and both $B_1(\omega)$ and $B_2(\omega)$ are regular
at $\omega=1$. At the zeros $\epsilon_0$ of $A(\epsilon)$, the singularity vanishes. For the corresponding
$\lambda=1/\epsilon_0-m$, both independent solutions are regular at $\omega=1$. Therefore solutions regular
both at $\omega=0$ and at $\omega=1$ can be constructed, and $\lambda$ is an eigenvalue for the particular
value $\epsilon_0$. In our case, these special points in the $(\epsilon, \, \lambda)$-plane can be
determined as roots of polynomials in $\epsilon$ with integer coefficients; and therefore they can be
determined with unlimited numerical precision.

We base the calculation on the formal Taylor expansion about $\omega=1$:
\begin{equation} \label{Karneol}
\beta_\lambda^{(\epsilon)}(\omega) = \sum_{k=0}^{\infty} d_{k}  (1-\omega)^{k} \, .
\end{equation}
The ODE (\ref{Falkenauge}) yields the recursion relation
\begin{equation} \label{Katzenauge}
d_{k} = \frac {\epsilon k (k-2) \;d_{k-2}-[2 \lambda (1 + \epsilon k) + 3 \epsilon k (k - 1)]\, d_{k-1} }{2k
(1-\epsilon (\lambda+k))} ~~~\mbox{for } k \ge 1,
\end{equation}
where we take $d_{0}=1, \, d_{-1}=0$. For $\lambda$ as in (\ref{Jade}) and for $k=m$, the denominator
vanishes and for general $\epsilon$ the ansatz (\ref{Karneol}) breaks down, showing that
$\beta_\lambda^{(\epsilon)} (\omega)$ picks up the singular part with $A(\epsilon)\ne0$ in (\ref{Jaspis}).
However, if the numerator in Eq.~(\ref{Katzenauge}) vanishes, the solution stays regular.

In evaluating this condition, it is preferable to rewrite the
recursion relation (\ref{Katzenauge}) in terms
of
\begin{equation} \label{Kunzit}
P_{k} (\epsilon)  = \epsilon^{k} k ! \left(\prod^{k}_{j=1} (1-\epsilon (\lambda+j))\right) ~ d_{k} \, ,
\end{equation}
with $\lambda={1}/{\epsilon}-m$ inserted. This yields
\begin{eqnarray} \label{Labradorit}
P_{k} (\epsilon)
& = & \epsilon^{4} \frac {k}{2} (k-1) (k-2) (m-k+1)\, P_{k-2}(\epsilon) \nonumber \\
&& - \left [ 1+\epsilon (k-m)+\epsilon^{2} \frac {k}{2}\, (3 \,k-2\, m - 3) \right ] P_{k-1} (\epsilon) \,
\end{eqnarray}
with $P_{-1}(\epsilon)=0$ and $P_0(\epsilon)=1$. $P_{k} (\epsilon)$ is a polynomial in $\epsilon$ of degree
$2 k$, with integer coefficients. Any real zero $\epsilon_{0}$ of $P_{m} (\epsilon)$ yields an eigenvalue
$\lambda = {1}/{\epsilon_0}- m$.

\begin{figure}[h]
\centering
\includegraphics[width=10cm]{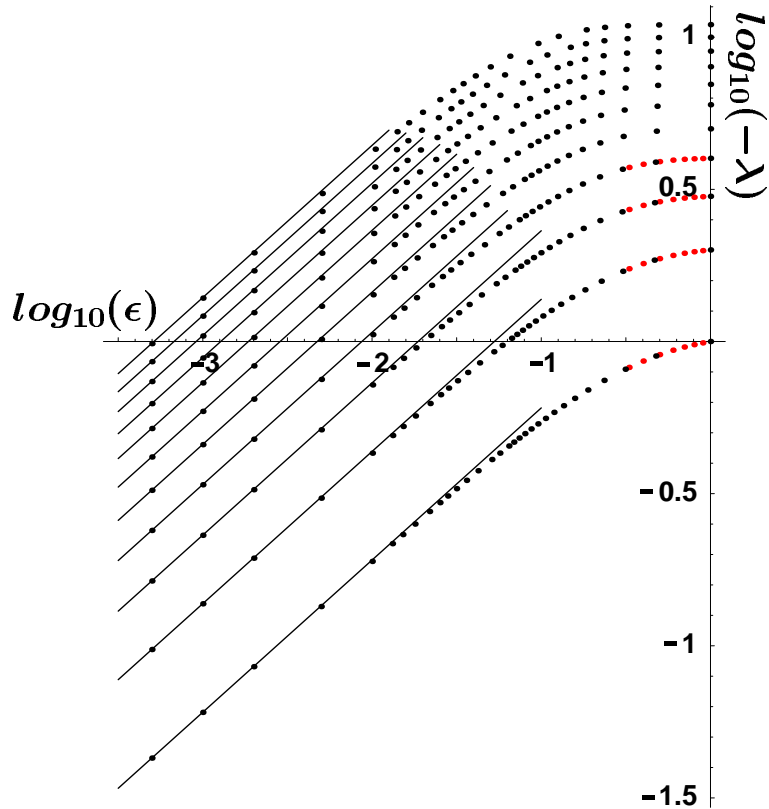}
\caption{Double logarithmic plot of the eigenvalues $\lambda_n(\epsilon)$, $n=1,\ldots,11$ (from bottom to
top) as a function of $\epsilon$ in the range $5\cdot10^{-4}\le\epsilon\le1$. The black dots indicate
discrete eigenvalues of the form (\ref{Jade}), gray (online: red) dots give eigenvalues calculated from the
Wronskian. The lines are interpolations of the form (\ref{Lapislazuli}). The coefficients $\alpha_n$
determined from these lines can be found in Table~\ref{table}.}\label{fig1}
\end{figure}

Using this approach we determined the $11$ largest negative eigenvalues for values of $m$ ranging from $m=2$
to $m=2000$, the latter value corresponding to $\epsilon \approx 5 \cdot 10^{-4}$. We note that the higher
coefficients of $P_{m} (\epsilon)$ with increasing $m$ become extremely large, but as function of
$\epsilon$, $P_{m} (\epsilon)$ oscillates around zero with an amplitude that for large $m$ becomes extremely
small in the range of interest $0<m- {1}/{\epsilon} ={\cal O} (1)$. For $m=2000$ this amplitude is of the
order $10^{-1000}$. Nevertheless the zeros of $P_{m} (\epsilon)$ can be determined with arbitrary precision
since the coefficients of $P_{m} (\epsilon)$ are integers, exactly determined from the recursion relation
(\ref{Labradorit}). Our results are shown in Fig.~\ref{fig1} in a double logarithmic plot. For $n$ fixed,
$|\lambda_{n} (\epsilon)|$ is seen to decrease with decreasing $\epsilon$. For $\epsilon\downarrow0$, the
results suggest the behavior
\begin{equation} \label{Lapislazuli}
\lambda_{n} (\epsilon) = - \alpha_{n} ~\epsilon^{1/2}~ \Big(1 + {\cal O} \,(\epsilon)\Big),
\end{equation}
reflecting the fact that the terms in ${\cal L}_\epsilon$ (\ref{Zirkonia}) that involve second
derivatives, for $\epsilon \downarrow 0$
play the role of singular perturbations. The coefficients
$\alpha_{n}$ extracted from our data are collected in the middle column of Table~\ref{table}. In the next
section we consider the limit $\epsilon \downarrow 0$ more closely.

\begin{table}
\centering
\begin{tabular}{|r|c|c|}
n & $\alpha_n$ from Section~\ref{negative eigenvalues} & $\alpha_n$ from Section~\ref{the limit} \\
\hline 1 & ~1.9131 & ~1.91 \\
2 & ~4.3516 & ~4.35 \\
3 & ~7.3107 & ~7.31 \\
4 & 10.7188 & 10.72 \\
5 & 14.5249 & 14.5~ \\
6 & 18.6916 & \\
7 & 23.1907 & \\
8 & 27.9992 & \\
9 & 33.0985 & \\
10 & 38.4732 & \\
11 & 44.1103 & \\
\hline
\end{tabular}
\caption{The coefficients $\alpha_n$ for $n=1,\ldots,11$ in the eigenvalue presentation
$\lambda_n(\epsilon)=-\alpha_n~\epsilon^{1/2}~(1+{\cal O}(\epsilon))$ from Eq.~(\ref{Lapislazuli}). The
middle column is extrapolated from Section~\ref{negative eigenvalues} and
Figure~\ref{fig1}. The right
column is from section~\ref{the limit}} \label{table}
\end{table}


\section{The eigenvalues for $\epsilon \downarrow 0$}\label{eps0}

\subsection{Eigenvalues with low indices}\label{the limit}

According to Eq.~(\ref{Granat}), the general solution of the ODE (\ref{Falkenauge}) at $\omega=1$ develops a
singularity that becomes arbitrarily weak if $\epsilon$ tends to zero. As a consequence, representing the
special solution $\beta_{reg} (\omega)$ regular at $\omega=0$ as a power series in $\epsilon$ and imposing
regularity at $\omega=1$ does not put any constraint on the eigenvalue $\lambda$. To show this we write
$\beta_{reg} (\omega)$ in the form
\begin{equation} \label{Larimar}
\beta_{reg} (\omega) = e^{\lambda \omega} \sum_{j=0}^\infty \epsilon^j f_j (\omega) \, ,
\end{equation}
imposing the normalization condition
\begin{displaymath}
\beta_{reg} (0) = 1 \, ,
\end{displaymath}
i.e.
\begin{equation}\label{Malachit}
f_0 (0) = 1, \quad f_j (0) = 0, \quad j \ge 1 \,.
\end{equation}
Substituting this ansatz into Eq.~(\ref{Falkenauge}) we find
\begin{equation} \label{Moldarit}
f_0 (\omega) \equiv 1
\end{equation}
\begin{eqnarray} \label{Mondstein}
f_j (\omega) &=& - \frac {1}{2} (1-\omega^2)\ \omega \ \partial_\omega f_{j-1} (\omega) + \lambda \omega^3
f_{j-1} (\omega)
\nonumber \\
&& + \frac{\lambda}{2} \int \limits_0^\omega dx \ (1+\lambda x - 3 x^2 + \lambda x^3)�\ f_{j-1} (x)
~~~\mbox{for } j \geq 1 \, .
\end{eqnarray}
Evidently $f_j (\omega)$ is a polynomial in $\omega$ of degree $4 j$. Thus $\beta_{reg} (\omega)$,
Eq.~(\ref{Larimar}), evaluated to arbitrary finite order in $\epsilon$, is analytic at $\omega=1$ for
arbitrary $\lambda$. This breakdown of an approach based on a simple expansion in the regularizing parameter
$\epsilon$ is a well known feature of such moving boundary problems, both in steady state (see for instance
\cite{Combescotetal86}) and linear stability analysis (see for instance \cite{Tanveer87c}).

For later use we note the expression for the low order terms:
\begin{eqnarray} \label{Mookait}
f_1 (\omega) & = & \frac {\lambda \omega}{2}   \left (1 + \omega^2 \right) + \frac {\lambda^2 \omega^2}{4}
\left(1 + \frac {1}{2} \omega^2 \right),
\nonumber\\
f_2 (\omega) & = & - \frac {\lambda \omega}{4} \left (1+2 \omega^2-3 \omega^4 \right) + \frac {\lambda^2
\omega^2}{8} \left (- 1 + 3 \omega^2 + 5 \omega^4 \right)
\nonumber\\
&&  + \frac {\lambda^3 \omega^3}{16}
\left(2 + \frac {23}{5} \omega^3 + \frac {15}{7} \omega^4 \right)�+ \frac {\lambda^4 \omega^4}{32} \ \left
(1 + \omega^2 + \frac {\omega^4}{4} \right) \,.
\end{eqnarray}
We also note the singular solution of Eq.~(\ref {Falkenauge}), resulting from a WKB-analysis
\begin{equation} \label{Morganit}
\beta_{sing} (\omega) = \left (1 - \frac {1}{\omega^2}\right )^{1/\epsilon} \left (\frac
{1+\omega}{1-\omega}\right)^\lambda e^{-\lambda \omega} \left( 1 + {\cal O} (\epsilon) \right ) \, .
\end{equation}
This solution picks up the singularities both at $\omega=0$ and at $\omega=\pm1$.

The above analysis demonstrates that for $\epsilon \downarrow 0$ the eigenvalues $\lambda$ can be determined
only by an analysis going beyond all orders in $\epsilon$. Indeed, taking into account the behavior $\lambda
\sim \epsilon^{1/2}$ suggested by section~\ref{negative eigenvalues}, it is evident that the individual
terms of the asymptotic expansion (\ref {Larimar}) become of order $1$ for $|\omega| \sim \epsilon^{-1/2}$
and consequently the expansion becomes invalid. Since, however, the eigenfunctions
$\beta_\lambda^{(\epsilon)}(\omega)$ by definition are holomorphic in the right half $\omega$ plane, this
suggests to analyze the region $|\omega| \sim \epsilon^{-1/2}$ more closely.

We therefore introduce the scaled variable
\begin{equation}\label{Nephrit}
\gamma = \frac {1}{\epsilon^{1/2} \, \omega}
\end{equation}
and the notation
\begin{eqnarray} \label{Obsidian}
\lambda &=& - \alpha ~ \epsilon^{1/2} \, , \\
\label {Onyx} \beta_\lambda^{(\epsilon)} \left(\omega (\gamma)\right) &=& g (\gamma) \, .
\end{eqnarray}
With these substitutions, the operator $L (\epsilon, \lambda)$ from (\ref{Zirkonia}) takes the form
\begin{eqnarray}
\label{Opal} L (\epsilon, \lambda) & = & \frac {\epsilon^{1/2}} {2} ~\gamma~
\Big(L_0 (\alpha) - \epsilon \, L_1 (\alpha) \Big) \, , \\
\label{Peridot} L_0 (\alpha) & = & \partial^2_\gamma + \left(2 \gamma -
\frac1\gamma \right) \partial_\gamma -
\frac {2 \alpha}{\gamma} \, , \\
\label{Pietersit}
L_1 (\alpha) & = & \gamma^2 \partial^2_{\gamma} + \left(\gamma-2 \alpha \right)
\partial_\gamma + \frac {2 \alpha}{\gamma} \, .
\end{eqnarray}

To leading order in $\epsilon$ we have to discuss the ODE
\begin {equation} \label{Prasem}
L_0 (\alpha) ~ g (\gamma) = 0 \, .
\end {equation}
An asymptotic expansion in powers of $\frac {1}{\gamma}$ yields the solution
\begin {equation}\label{Pyrit}
g_1 (\gamma) \sim 1 - \frac {\alpha}{\gamma} + \frac {\alpha^2}{2 \gamma^2}
- \frac {\alpha}{2 \gamma^3} \left(1 +
\frac {\alpha^2}{3} \right) + {\cal O} \left(\frac {1}{\gamma^4}\right) \, .
\end {equation}
The second solution, found by balancing the terms $\partial^2_\gamma$ and
$(2 \gamma - \frac {1}{\gamma})
\,\partial_\gamma$, takes the form
\begin {equation} \label {Rauchquarz}
g_2 (\gamma) \sim e^{-\gamma^2} \left(1 + {\cal O}\,
\left (\frac {1}{\gamma}\right) \right) \, .
\end {equation}
It is easily seen that in the range
\begin {displaymath}
\epsilon^{-1/2} \gg |\gamma|
\gg 1, \quad \mbox{i.e.,} \quad 1 \ll |\omega| \ll \epsilon^{-1/2} \, ,
\end {displaymath}
$g_1 (\gamma)$ matches the regular solution $\beta_{reg} (\omega)$ given in (\ref{Larimar}) and
(\ref{Mookait}), whereas $g_2 (\gamma)$ matches the singular solution $\beta_{sing} (\omega)$ given in
(\ref{Morganit}).

Evidently $g_2 (\gamma)$ dominates over $g_1 (\gamma)$ in the two sectors $\frac {\pi}{4} < |\arg \gamma| <
\frac {3}{4} \pi$ for both signs of $\arg\gamma$. We search for eigenfunctions $\beta (\omega)$ matching
$g_1 (\gamma)$ in the complete range $|\arg \gamma| < \frac {\pi}{2}$. Now it is well known that starting
deep in the region $\frac {\pi}{4} < \arg \gamma < \frac {\pi}{2}$ with initial conditions taken from the
asymptotic expansion (\ref{Pyrit}) and integrating Eq.~(\ref{Prasem}) down into the region $- \frac {\pi}{2}
< \arg \gamma < - \frac {\pi}{4}$, we in general will pick up a dominant contribution proportional to $g_2
(\gamma)$. If, however, $g (\gamma) \in \mathbb{R}$ for $\gamma \in \mathbb{R}$, the Schwarz reflection
principle guarantees the absence of such a contribution. Equivalently, we may state that by definition the
eigenfunctions $\beta (\omega)$ are real for $\omega \in (0, 1)$, and that the absence of the singularity
induced cut for $\omega > 1$ implies $\beta (\omega) \in \mathbb{R}$ for $\omega > 0$. Thus the eigenvalues
$\lambda_n^{(\epsilon)} = - \alpha_n \, \epsilon^{1/2} (1 + {\cal O} (\epsilon))$ are selected by imposing
the constraint $g (\gamma) \in \mathbb{R}$ for $\gamma > 0$.

\begin{figure}[h]
\begin{center}
\includegraphics[width=8cm,angle=0]{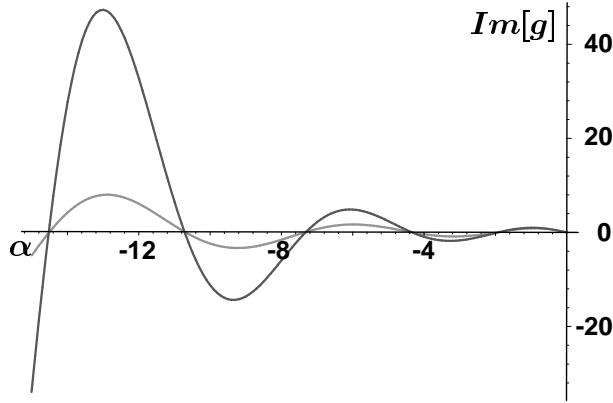}
\caption{$Im ~g(b)$ as a function of $\alpha$ for
two different choices of the starting point $\gamma_0=b+iL$,
$b=0.1$ and 0.3; the zeros of the function determine $\alpha_n$.}\label{fig2}
\end{center}
\end{figure}

To evaluate this criterion we started at points $\gamma_0 =
b+i L$, $b = 0.1,~ 0.2, ~0.3$, $L = 20,~ 25,~ 30$
and integrated Eq.~(\ref{Prasem}) down to the real axis along lines $b={\rm const}$. The results for the
first $5$ positive zeros $\alpha_n$ of ${\rm Im}[g (b)]$ are given in the right column of Table~\ref{table}.
They clearly are consistent with the asymptotic results found in subsection~\ref{negative eigenvalues} and
presented in the middle column of the same table. We did check that the quoted values to the precision given
are insensitive to the starting value $\gamma_0$
as long as $L$ is chosen sufficiently large and $b > 0$ is
not too large or too small. The former causes ${\rm Im}[g (b)]$
to be rather small and zeros are harder to
detect, while the latter causes numerical inaccuracies
due to the singularity of $L_0 (\alpha)$ at $\gamma =
0$. Fig.~\ref{fig2} shows that with different $b$, while
${\rm Im}[g (b)]$ itself is different, the zeros
match as expected from theory.

\subsection{Eigenvalues with large indices $n$}

Both the approach of the last section and that of Sect.~\ref{negative eigenvalues} within reasonable
numerical effort yield the asymptotic coefficients $\alpha_n$ only for the first few eigenvalues, $\alpha_n
= {\cal O}(1)$. However, in the complementary region $\alpha_n \gg 1$, a fully analytical analysis is
possible, similar to that performed earlier in the surface tension selection problem for steady Hele-Shaw
fingers \cite{Combescotetal86,Tanveer87b,DorseyMartin}. We first introduce a Liouville transformation to
eliminate the first order derivative in the operator $L_0(\alpha)$. With
\begin{equation}
\label{51} g(\gamma)=\gamma^{1/2}~e^{-\gamma^2/2}~h(\gamma),
\end{equation}
Eq.~(\ref{Prasem}) takes the form
\begin{equation}
\label{52}
\partial_\gamma^2 h-\left(\gamma^2+\frac{2\alpha}\gamma + \frac3{4\gamma^2}
\right) h=0.
\end{equation}
We now rescale $\gamma$ according to
\begin{equation}
\label{53}
\gamma=\alpha^{1/3}\chi,~~~h(\gamma(\chi))=q(\chi)
\end{equation}
to find
\begin{equation}
\label{54}
\partial^2_\chi q-\alpha^{4/3}
\left(\chi^2+\frac2\chi+\frac 3 {4\alpha^{4/3}\chi^2}\right)q=0.
\end{equation}

For $\alpha\to\infty$, a WKB-analysis yields the asymptotic relation
\begin{eqnarray}
\label{55}
 q(\chi)&\sim&C_1~q_1(\chi)+C_2~q_2(\chi),
\\
\label{56} q_{1,2}(\chi)&=& Q^{-1/4}(\chi)~\exp\left[\pm\alpha^{2/3}\int_0^\chi
d\chi'~Q^{1/2}(\chi')\right],
\end{eqnarray}
where
\begin{equation}
\label{57}
Q(\chi)=\chi^2+\frac2\chi.
\end{equation}
$q_2(\chi)$ dominates in the sectors $\frac\pi4 < |\arg \chi| < \frac\pi2$, but is subdominant on the real
axis. Recalling Eqs.~(\ref{51}) and (\ref{53}),
it is easily seen that for large $|\chi|$, $q_1(\chi)$
yields $g_1(\gamma)$, Eq.~(\ref{Pyrit}),
whereas $q_2(\chi)$ yields $g_2(\gamma)$, Eq.~(\ref{Rauchquarz}). Since
the eigenfunctions in the limit $\epsilon \downarrow0$
in all the right half plane ${\rm Re}[\gamma]>0$ must
reduce to $g_1(\gamma)$, this implies that $C_1$ is real whereas
$C_2$ has to vanish in both sectors
$\frac\pi4 < |\arg \chi| < \frac\pi2$.

Now the WKB-analysis breaks down at the turning points
$\chi_s$ defined by $Q(\chi_s)=0$, which in the
relevant region ${\rm Re}[\chi] >0$ yields the two solutions
\begin{equation}
\label{58} \chi_s=2^{1/3}~e^{\pm i\pi/3}.
\end{equation}
As is well known in asymptotics \cite{Olver} the coefficients in the asymptotic relation (\ref{55}) may jump
when $\chi$ crosses a Stokes line emerging from a turning point. Thus a vanishing $C_2$ in the sector
$\frac\pi4 < \arg \chi < \frac\pi2$ does not imply that $C_2$ vanishes on the positive real axis as well.
Since, however, the eigenfunctions are real on the real axis, we get the relation
\begin{equation}
\label{59} {\rm Im}[C_2]=0 ~~~\mbox{for }\chi\in \mathbb{R}_+
\end{equation}
as a necessary condition fixing the eigenvalues. To evaluate this condition we determine the jump of $C_2$
by analyzing the neighborhood of the turning point $\chi_s=2^{1/3} e^{+i \pi/3}$.

Expanding $Q(\chi)$ about $\chi_s$,
\begin{eqnarray}
\label{60}
Q(\chi)&=&Q'(\chi_s)~\big(\chi-\chi_s\big)+{\cal O}\big(\chi-\chi_s\big)^2,
\\
\label{61}
Q'(\chi_s)&=&\frac 6 {2^{2/3}}~e^{i\pi/3}
\end{eqnarray}
and rescaling $\chi-\chi_s$ according to
\begin{eqnarray}
\label{62}
\xi&=&\alpha^{4/9}\left(\frac 6 {2^{2/3}}\right)^{1/3}~e^{-i5\pi/9}~\big(\chi-\chi_s\big),
\\
\label{63}
q(\chi(\xi))&=&m(\xi),
\end{eqnarray}
we find that Eq.~(\ref{54}) to leading order in $\alpha$ reduces to the Airy equation
\begin{equation}
\label{64}
\left(\partial_\xi^2-\xi\right) m(\xi)=0.
\end{equation}
The phase in Eq.~(\ref{62}) has been chosen such that for $\xi > 0$ we enter the region where $C_2=0$ in the
asymptotic relation (\ref{55}). This analysis is valid for $|\chi-\chi_s|\gg\alpha^{-4/9}$, and using
Eq.~(\ref{60}), we find for small $|\chi-\chi_s|$ along the line $\xi>0$
\begin{equation}
\label{65} q(\chi)\sim a~\xi^{-1/4}~\exp\left[-~\frac23\xi^{3/2}\right],
\end{equation}
where the constant $a$ is given by
\begin{equation}
\label{66} a=\alpha^{1/9}~C_1~\left(\frac{2^{2/3}}6\right)^{1/6}~e^{-i2\pi/9}~
\exp\left[\alpha^{2/3}\int_0^{\chi_s}d\chi'\sqrt{Q(\chi')}\right].
\end{equation}
This result is valid for $\alpha^{4/9}\gg\xi\gg1$, where it must match the solution of the Airy equation
(\ref{64}). This yields
\begin{equation}
\label{67} m(\xi)=2\sqrt{\pi}~a~{\rm Ai}(\xi).
\end{equation}
Continuing this result clockwise around the turning point to negative $\xi$, we find
\begin{equation}
\label{68} m(\xi)\sim a~|\xi|^{-1/4}~e^{i\pi/4}~\left(\exp\left[-~\frac23~i|\xi|^{3/2}\right]
-i~\exp\left[\frac23~i|\xi|^{3/2}\right]\right)~~~\mbox{for }-\xi\gg1.
\end{equation}

\begin{figure}
\begin{center}
\includegraphics[width=10cm]{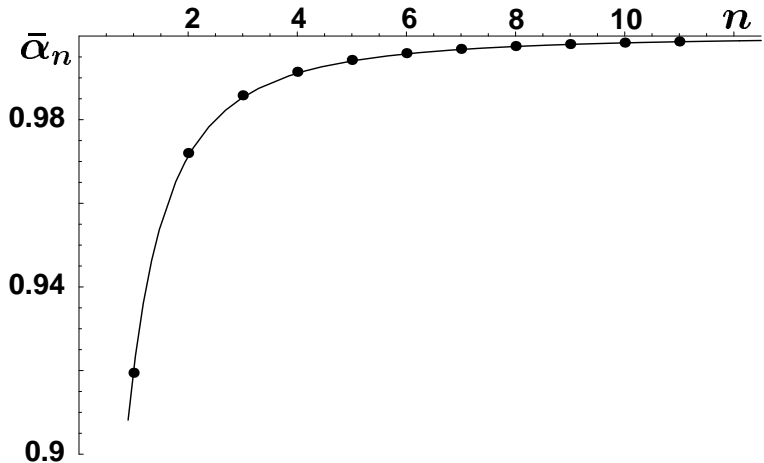}
\caption{$\bar\alpha_n$ from Eq.~(\ref{bar alpha}) as a function of $n$, together with the interpolating
formula (\ref{inter}). The plot demonstrates the accuracy of the asymptote~(\ref{73}) even for small
$n$.}\label{fig3}
\end{center}
\end{figure}

For $\alpha^{4/9}\gg-\xi\gg1$, this again has to match with Eq.~(\ref{55}), which in this region reduces to
\begin{eqnarray}
\label{69} q(\chi)&\sim& a~|\xi|^{-1/4}~e^{i\pi/4}~\left(\exp\left[-\frac23~i~|\xi|^{3/2}\right]\right.
\\
&&\qquad\qquad\left.+\frac{C_2}{C_1}~\exp\left[-2\alpha^{2/3}\int_0^{\chi_s}d\chi'\sqrt{Q(\chi')}\right]
~\exp\left[\frac23~i~|\xi|^{3/2}\right]\right). \nonumber
\end{eqnarray}
We thus find
\begin{equation}
\label{70} C_2=-iC_1~\exp\left[2\alpha^{2/3}\int_0^{\chi_s}d\chi'\sqrt{Q(\chi')}\right].
\end{equation}
Since both $C_1$ and $C_2$ have to be real, this leads to the eigenvalue condition
\begin{equation}
\label{71} {\rm Im}
\left[2\alpha^{2/3}\int_0^{\chi_s}d\chi'\sqrt{Q(\chi')}\right]=\left(n+\frac12\right)\pi,~~~n\in\mathbb{Z}.
\end{equation}
The integral is easily evaluated to yield
\begin{eqnarray}
\label{72} \int_0^{\chi_s}d\chi'\sqrt{Q(\chi')}&=&2^{2/3}~e^{i\pi/6}\int_0^1 ds\sqrt{\frac1s-s^2}
\nonumber\\
&=& 2^{2/3}~\frac{\sqrt{\pi}}8~\frac{\Gamma\left(\frac16\right)}{\Gamma\left(\frac23\right)} ~\big(\sqrt 3
+i\big).
\end{eqnarray}
Eq.~(\ref{71}) yields the final result
\begin{eqnarray}\label{alpha_new}
\label{73}
\alpha_n&=&4\left(\frac{\Gamma\left(\frac23\right)}{\Gamma\left(\frac16\right)}~\sqrt\pi
~\left(n+\frac12\right)\right)^{3/2} \\
\label{74} &=& 1.13254\ldots~\left(n+\frac12\right)^{3/2}~~~\mbox{for }n\in\mathbb{N}_0.
\end{eqnarray}
By construction this result is valid in the limit $\alpha_n \to \infty$, i.e., for $n \to \infty$. However,
it turns out to be a good approximation for small $n$ as well. This is illustrated in Fig.~\ref{fig3} where
we plot
\begin{equation}\label{bar alpha}
\bar\alpha_n=\frac{\alpha_n}4~ \left(\frac{\Gamma\left(\frac23\right)}{\Gamma\left(\frac16\right)}~
\sqrt\pi~\left(n+\frac12\right)\right)^{-3/2},
\end{equation}
with $\alpha_n,~n=1,\ldots,11$ taken from the second line of Table~\ref{table}. Even for $n=2$ the
asymptotic result (\ref{73}) differs from the true eigenvalue by less than 3 \%. The interpolating curve
included in Fig.~\ref{fig3} is given by
\begin{equation}\label{inter}
\bar{\alpha}_n=1-0.18~\left(n+\frac12\right)^{-2},
\end{equation}
indicating that the asymptotic limit (\ref{73}) is rapidly approached.

\section{Behavior of low order eigenfunctions for $|\omega| \leq 1$}\label{EigenFunc}

As is evident from Eq.~(\ref{Granat}), for $\omega \to - 1$ the eigenfunctions are singular,
\begin{displaymath}
\beta_{\lambda}^{(\epsilon)} (\omega) \sim (1 + \omega)^{\frac {1}{\epsilon}+ \lambda}.
\end{displaymath}
For small $\epsilon$ and $| \lambda | \ll 1/\epsilon$, this singularity will show up only in derivatives of
high order, and the Taylor expansion (\ref{Holzstein}) will converge in the whole physical domain
$|\omega|\le1$. We therefore can use this expansion together with the recursion relation (\ref{Howlith}),
(\ref{Zinnstein}) to evaluate $\beta_{\lambda}^{(\epsilon)}(\omega)$ for the first few eigenvalues. Outside
some neighborhood of $\omega = - 1$ these functions are expected to govern the large time behavior of
analytic perturbations.

\begin{figure}
\begin{center}
\includegraphics[width=6cm]{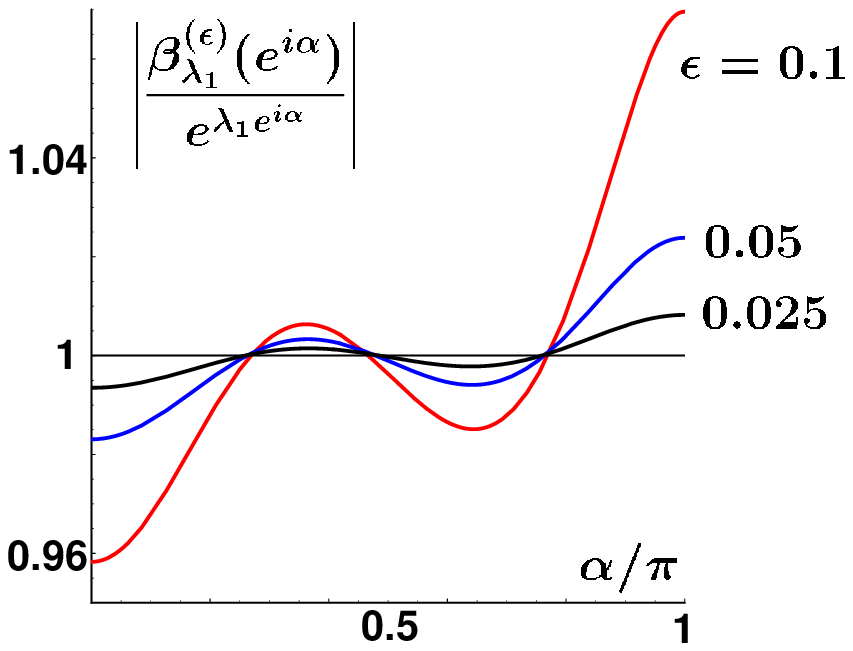}
\includegraphics[width=6cm]{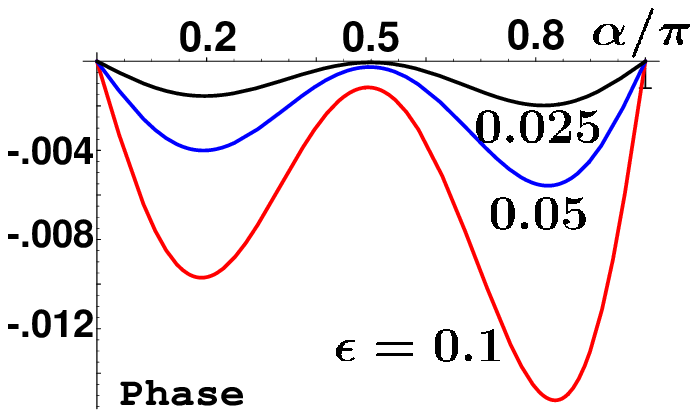}
\caption{The ratio ${\cal F}=\beta^{(\epsilon)}_{\lambda_{1}} (\omega)/ e^{\lambda_1 \omega}$ evaluated on
the unit circle $\omega = e^{i \alpha}$ for $\epsilon=0.1,~0.05,~0.025$ as indicated. a) Absolute value
$|{\cal F}|$ and b) argument ${\rm Arg}({\cal F})/\pi$ as a function of $\alpha/\pi$.}\label{fig4}
\end{center}
\end{figure}

\begin{figure}
\begin{center}
\includegraphics[width=6cm]{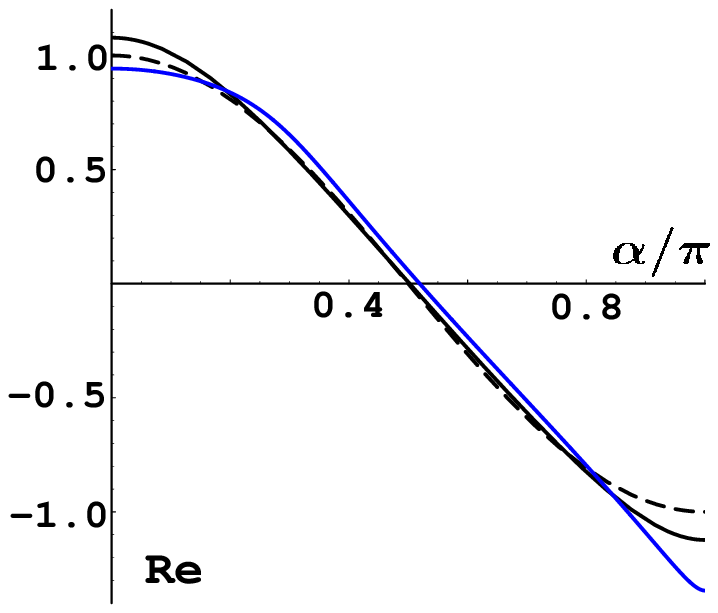}
\includegraphics[width=6cm]{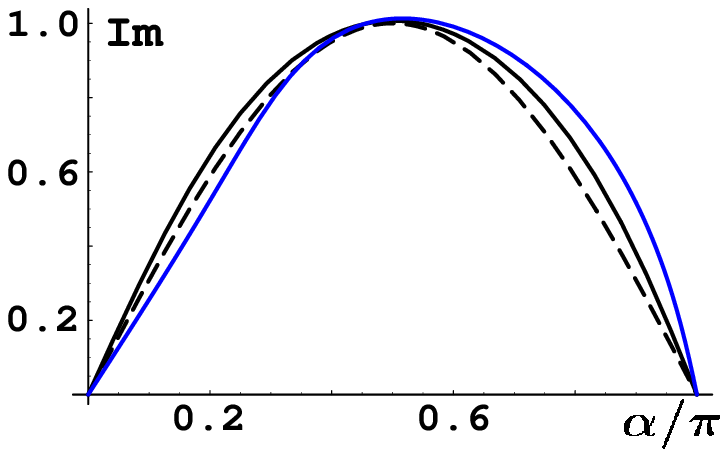}
\caption{Real part (a) and imaginary part (b) of the function ${\cal
G}=\ln\left[\beta_\lambda^{(\epsilon)}(e^{i\alpha})\right]/\lambda$ for $\epsilon=0.1$ for $\lambda_1$
(black solid line) and $\lambda_6$ (blue or grey solid line) as a function of $\alpha/\pi$. The dashed lines
show $\cos\alpha$ in (a) or $\sin\alpha$ in (b) for comparison.}\label{fig5}
\end{center}
\end{figure}

>From the discussion of $\beta_{reg}(\omega)$ in Subsect.~\ref{the limit} we expect that in the physical
domain $|\omega|\le1$ the eigenfunctions for small $\epsilon$ are well approximated by simple exponentials.
This indeed is true as illustrated in Fig.~\ref{fig4}. Panel~\ref{fig4}(a) shows the absolute value $|
\beta^{(\epsilon)}_{\lambda_{1}} (\omega)\, e^{- \lambda_1 \omega} |$ evaluated on the unit circle $\omega =
e^{i \alpha}$ for a decreasing set of values of $\epsilon$. Evidently this ratio tends to 1 and is not very
different from 1 even for $\epsilon = 0.1$. Panel~\ref{fig4}(b) shows the corresponding phase difference
which is found to be very small and to tend to zero with decreasing $\epsilon$. All curves in
Fig.~\ref{fig4} are well represented by the expansion (\ref{Larimar}) evaluated up to order $\epsilon^2$. As
expected, for fixed $\epsilon$ the approximation
\begin{equation}\label{EF}
\beta_{\lambda}^{(\epsilon)} (e^{i\, \alpha}) \sim \exp \, (\lambda \cos \alpha + i \lambda \, \sin \alpha)
\end{equation}
becomes worse with increasing $\lambda$. This is illustrated in Fig.~\ref{fig5} for $\epsilon = 0.1$,
$\lambda_{1} = - 0.5462$, $\lambda_{6} = - 4.6086$. Panel~\ref{fig5}(a) compares ${\rm Re} \left( \ln \,
\beta_{\lambda}^{(\epsilon)} (e^{ i \alpha}) \right) / \lambda$ with $\cos \alpha$. In the neighborhood of
$\alpha = \pi$ the deviation from $\cos \alpha$ increases with increasing $\lambda$, which is a consequence
of the singularity of $\omega = - 1$. But outside that range the approximation (\ref{EF}) is quite accurate
even for $\beta_{\lambda_6}^{(1)}$. Panel~\ref{fig5}(b) shows the corresponding results for ${\rm Im} \left(
\ln \, \beta_{\lambda}^{(\epsilon)} (e^{i \alpha}) \right) / \lambda$.

As pointed out in section~\ref{precise formulation}, we expect that the initial condition $\beta(\omega,0)$
in some neighborhood of $\omega=1$ can be expanded in terms of eigenfunctions as
\begin{equation}
\beta(\omega,0)=\sum_{k=0}^\infty g_n\,\beta_{\lambda_n}^{(\epsilon)}(\omega),
\end{equation}
and that the domain of convergence of this expansion increases with time to ultimately cover ${\cal
U}_\omega\setminus\{-1\}$. For small $\epsilon$ and not too large $|\lambda|$,
$\beta_\lambda^{(\epsilon)}(\omega)$ is well approximated by $e^{\lambda\omega}$, and all eigenvalues vanish
for $\epsilon\to0$. Thus for fixed $\omega$ and $\epsilon\to0$, all eigenfunctions tend to 1. This suggests
that for small $\epsilon$ many terms must contribute non-negligibly to the expansion of a generic initial
condition. Furthermore the phase of the coefficients $g_n$ must vary strongly. With decreasing $\epsilon$ it
will need more and larger coefficients to describe a non-trivial general initial condition. These large
coefficients imply large transient growth, because in the representation
\begin{equation}
\beta(\omega,t)=\sum_{k=0}^\infty g_n\,\beta_{\lambda_n}^{(\epsilon)}(\omega) e^{\lambda_n t} ,
\end{equation}
the factor $e^{\lambda_n t}$ is substantially different for each $n$ when $t=O (\epsilon^{-1/2}) $. The
large transient growth restricts the validity of linear stability analysis to a small ball of initial
conditions. This is confirmed in part II in the study of the initial value problem.



\section{Conclusion and outlook}\label{Concl}

In the introduction we raised the question: Can a kinetic undercooling boundary condition regularize the
evolution of a curved interface in a Laplacian growth model? This problem is motivated by the physics of
streamer discharges which determine the early evolution of sparks and lightning. We here gave a first answer
by analyzing the spectrum of linear perturbations of uniformly translating circles. We proved that for all
$\epsilon>0$ the spectrum is discrete and that all eigenvalues $\lambda_n$, except for $\lambda_0=0$, have a
negative real part. Thus any infinitesimal perturbation tends to a constant exponentially in time for large
time, and asymptotically the circular shape is recovered. For arbitrary regularization parameter
$\epsilon>0$ we found an infinite set of negative real eigenvalues $\lambda_n{(\epsilon)}$; they smoothly
continue the exact result $\lambda_n{(1)}=-n$, $n \in \mathbb{N}$, found previously for
$\epsilon=1$~\cite{eber07}.

In formulating the eigenvalue problem we had to allow for a singularity of the eigenfunctions at the point
$\omega=-1$ at the back of the circle. Therefore an expansion of a regular initial perturbation in terms of
eigenfunctions must break down in a neighborhood of $\omega=-1$. For $\epsilon=1$, it was
found~\cite{eber07} that the size of this neighborhood with increasing time $\tau$ decreases as $e^{-\tau}$,
and that all the structure of the initial perturbation is convected into this region. In part
II~\cite{partII} of this series of papers we will argue that such behavior is found for all $\epsilon>0$.

Our results suggest that in the framework of linear perturbation theory the evolution of a curved front can
be regularized by a kinetic undercooling boundary condition. Linear perturbation theory must break down in
the limit $\epsilon\downarrow0$. In our results this shows up in the asymptotic behavior of the eigenvalues
$\lambda_n{(\epsilon)} = -\alpha_n~\epsilon^{1/2}$, where $\alpha_n\sim {\rm const}~n^{3/2}$ for large $n$,
and in the associated behavior $\beta_{\lambda_n}^{(\epsilon)}(\omega)\to e^{\lambda_n\omega}\to1$ of the
eigenfunctions. This indicates that for small $\epsilon$, the eigenfunction expansion of even a very smooth
initial perturbation of small but finite amplitude will contain many terms with large coefficients.
Initially these terms almost compensate, but the balance is destroyed by the temporal evolution. Generically
this will lead to a strong transient growth of the perturbation which may drive the evolution into the
regime where nonlinear effects have to be included. Examples supporting this scenario will be given in part
II~\cite{partII}. For $\epsilon\ll1$ this mechanism can lead to an instability of the circular shape also
against quite small and smooth perturbations.
\\

{\bf Acknowledgements:} S. Tanveer acknowledges hospitality at CWI Amsterdam. Additional support was
provided by the U.S. National Science Foundation (DMS-0405837, DMS-0733778, DMS-0807266). F. Brau
acknowledges a grant of The Netherlands' Organization for Scientific Research NWO within the FOM/EW-program
"Dynamics of Patterns".

\newpage

\begin{appendix}

\section{Appendix A: Asymptotics of Taylor series coefficients and relation
to singularity at $\omega=\omega_s$}
\label{AppA}

Here we prove the following standard results for the sake of completeness.

\noindent{\bf Theorem:}
{\it Assume $f(\omega)$ is analytic in $|\omega| < R$, except for a
singularity at
$\omega = \omega_s \ne 0 $ (with $|\omega_s| < R$), where
$$f(\omega) = f_s (\omega) \left ( \omega - \omega_s \right )^{\alpha} +
f_a (\omega) ~~~{\rm as} ~~\omega \rightarrow \omega_s. $$ Here $\alpha$ is not a positive integer, and $f_s
$ and $f_a$ are locally analytic at $\omega=\omega_s$ with $f_s (\omega_s) \equiv C \ne 0$. Then the Taylor
series coefficient $b_k$ in the representation
$$ f(\omega) = \sum_{k=0}^\infty b_k \omega^k $$
has the leading order asymptotic behavior
$$ b_k \sim - \frac{C e^{i \pi \alpha}}{\pi} \sin (\pi \alpha)
\omega_s^{-k+\alpha} \frac{\Gamma (\alpha+1)}{k^{\alpha+1}}. $$}

\begin{figure}[h]
\begin{center}
\includegraphics[height=8cm,angle=0]{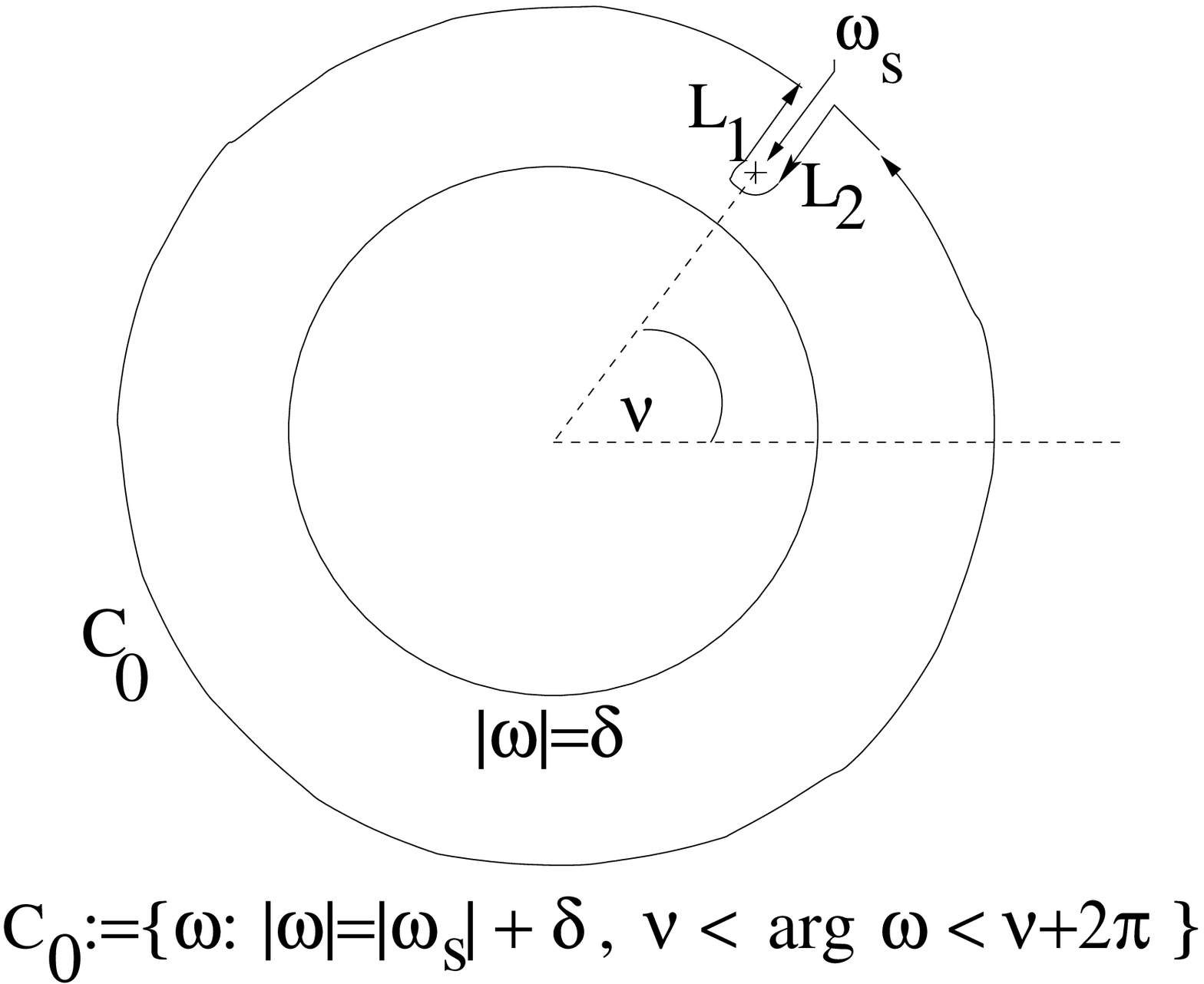}
\caption{Contour $\oint_{|\omega|=\delta} $ deformed to $\int_{C_0}+
\int_{L_1} + \int_{L_2} $ for evaluation of $b_k$.} \label{fig6}
\end{center}
\end{figure}

\noindent{\bf Proof:} We will first assume $\alpha > -1$.
We recall the contour integral representation of $b_k$
$$ b_k = \frac{1}{2 \pi i} \oint_{|\omega|=\delta}
\frac{f(\omega) }{\omega^{k+1} }~ d\omega ,
$$
where $\delta$ is chosen sufficiently small so that $|\omega|=\delta$ contains no singularity of $f(\omega)$
and $|\omega_s|+\delta < R$. We deform the contour into $\int_{C_0}+\int_{L_1}+\int_{L_2}$ as shown in
Figure \ref{fig6}. The contribution from $\int_{C_0}$ is easily bounded by $M \left [ |\omega_s| + \delta
\right ]^{-k} $, where $M = \sup_{|\omega|=|\omega_s|+\delta} |f(\omega)|$. Now consider the contribution
from $\int_{L_1}+\int_{L_2}$. It is convenient to introduce a change of variable
$$ \zeta = \log \frac{\omega}{\omega_s}.$$

Then it is readily checked that we have
\begin{equation}
\label{L1L2cont} \int_{L_1}+\int_{L_2} = \frac{\omega_s^{-k}}{2 \pi i} \left \{ \int_0^{\delta_1} - \int_{0
e^{2 i \pi}}^{\delta_1 e^{2i \pi}} \right \} f (\omega (\zeta)) e^{-k \zeta} d\zeta,
\end{equation}
where $\delta_1 = \log \left [ 1 + \frac{\delta}{|\omega_s|} \right ]$.
Noting that the contribution from $f_a $ cancels out between $L_1$ and
$L_2$, we obtain
\begin{equation*}
\int_{L_1}+\int_{L_2} = \frac{\omega_s^{-k+\alpha} }{ 2 \pi i} \left [ 1- e^{2 i \pi \alpha} \right ]
\int_0^{\delta_1} \left [ e^{\zeta} -1 \right ]^{\alpha} f_s (\omega (\zeta)) e^{-k \zeta} d\zeta.
\end{equation*}
We note that in the neighborhood of $\zeta=0$,
$$ \left [ e^{\zeta} - 1 \right ]^\alpha f_s (\omega (\zeta) )
\sim C \zeta^\alpha .$$ So, using Watson's Lemma, we obtain
$$
\int_{L_1}+\int_{L_2} \sim \frac{C}{2 \pi i}
\omega_s^{-k +\alpha} \left [
1-e^{2 \pi i \alpha}
\right ]  k^{-1-\alpha} \Gamma (1+\alpha), $$
from which the Lemma follows since the contribution from $\int_{C_0}$ is
evidently exponentially small relatively for large $k$.

If $\alpha < -1$, we consider
$n$-th iterated integral $I_n f $, where $[I_1 f] (\omega) = \int_0^\omega
f (\omega') d \omega' $, $I_2 f = I_1 \left ( I_1 f \right )$ and so on.
Then, it is clear that the singularity of $I_n f $ at
$\omega = \omega_s$ will be of the type
$(\omega-\omega_s)^{\alpha+n}$, and we can arrange $\alpha+n > -1 $.
The argument above can then be repeated for power series coefficients of
$I_n f$. The Theorem follows on differentiating $n$-times the power series of
$I_n f$.

\noindent{\bf Remark} {\it The previous theorem
remains valid even when $\alpha $ is
a negative integer, provided the product of
$\sin (\pi \alpha) \Gamma (1+\alpha) $ is replaced by its finite nonzero limit
$\lim_{\alpha \rightarrow -n}$. The validity of the result is easily
checked from Taylor expansion of
$\frac{1}{(\omega-\omega_s)}$ and its derivatives
in terms of a geometric series and its derivatives. Further, since
the asymptotic result relies on
Watson's Lemma, the condition on analyticity of $f_s$ at $\omega=\omega_s$
can be weakened to
$f_s (\omega) = C +
O \left ( ( \omega-\omega_s )^{\beta} \right ) $ for $\beta > 0$.}

\noindent{\bf Theorem:} {\it If the $k$-th Taylor series coefficient
$b_k$ of an
analytic function $f$ at $\omega=0$ satisfies the following
\begin{equation}
\label{bkexp}
b_k \sim - \frac{C}{\pi} e^{i \pi \alpha} \sin (\pi \alpha)
\omega_s^{-k+\alpha}
\frac{\Gamma (\alpha+1)}{k^{\alpha+1}} \left ( 1+ O(1/k) \right )
\end{equation}
for non-integral $\alpha$, then
\begin{equation}
\label{bkexp1} f(\omega) \sim C (\omega-\omega_s)^{\alpha} + f_a (\omega),
\end{equation}
where $f_a (\omega)$ is regular at $\omega=\omega_s$.
The same conclusion is valid for $\alpha = -n$, a negative integer, provided
$\Gamma (1+\alpha)
\sin (\pi \alpha)$ is replaced by its limit as $\alpha \rightarrow -n$,
On the otherhand, if
$$ b_k \sim - C \omega_s^{-k} k^{-1}
\left (1 + O(1/k) \right ), $$
then
$$ f(\omega) \sim C \log \left (1- \frac{\omega}{\omega_s} \right ). $$
}

\noindent{\bf Proof}.
We will assume for now that $\alpha \in (-1, 0)$.
Then using previous Theorem to determine Taylor series coefficient
of $C (\omega-\omega_s)^{\alpha}$, it follows on subtraction that
$$g(\omega) \equiv f(\omega) - C (\omega-\omega_s)^{\alpha}
= \sum_{k=0}^\infty b_k \omega^k $$
will have Taylor series coefficient $b_k \sim ~{\rm Const.}~ \omega_s^{-k}
k^{-\alpha-2}$ for large $k$ implying that the series is absolutely
convergent on
$|\omega|=|\omega_s|$. In particular, $g$ is
continuous at $\omega=\omega_s$, which proves the theorem for
$\alpha \in (-1, 0) $.
If $\alpha$ takes on
other ranges of values, we obtain this result by either $n$-times
iterative integration from the origin of the power series for $f$,
or differenting it
$n$ times so as to ensure that for non-integral $\alpha$,
$\alpha +n \in (-1,0)$ or $\alpha-n \in (-1, 0)$.
The result quoted in the Theorem
follows by noting that the asymptotics is differentiable
as it is valid for $\omega-\omega_s$ in a complex sector.
The second result follows from noting the explicit Taylor expansion of
$\log \left (1 - \frac{\omega}{\omega_s} \right )$ and noting
that the difference has a series that is absolutely summable and hence
the remainder is continuous at $\omega=\omega_s$.
By using explicit derivatives or integrals of
Geometric series, the conclusion (\ref{bkexp1}) holds for
for negative integer $\alpha=-n$ as well.

\end{appendix}

\end{document}